\documentclass[nofootinbib, preprint, aps, prd]{revtex4-1}

\usepackage{graphicx}
\usepackage{natbib}

\newcommand{\beq}{\begin{equation}}
\newcommand{\beqa}{\begin{eqnarray}}
\newcommand{\eeq}{\end{equation}}
\newcommand{\eeqa}{\end{eqnarray}}

\newcommand{\simgt}{\lower.5ex\hbox{$\; \buildrel > \over \sim \;$}}
\newcommand{\simlt}{\lower.5ex\hbox{$\; \buildrel < \over \sim \;$}}

\newcommand{\bd}[1]{\mbox{\boldmath $#1$}}

\usepackage[usenames,dvipsnames]{xcolor}

\begin{document}
\title{
Cross-Correlation of Cosmic Shear and Extragalactic Gamma-ray Background: Constraints on the Dark Matter Annihilation Cross-Section
}

\author{Masato Shirasaki}
\affiliation{Department of Physics, University of Tokyo, Tokyo 113-0033, Japan}
\email{masato.shirasaki@utap.phys.s.u-tokyo.ac.jp}

\author{Shunsaku Horiuchi}
\affiliation{Center for Cosmology, Department of Physics and Astronomy, 4129 Frederick Reines Hall, University of California, Irvine, CA 92697-4575}
\email{s.horiuchi@uci.edu}

\author{Naoki Yoshida}
\affiliation{
Department of Physics, University of Tokyo, Tokyo 113-0033, Japan\\
Kavli Institute for the Physics and Mathematics of the Universe (WPI),
University of Tokyo, Kashiwa, Chiba 277-8583, Japan
}

\begin{abstract}

We present the first measurement of the cross-correlation of weak gravitational lensing 
and the extragalactic $\gamma$-ray background emission using data from the Canada-France-Hawaii 
Lensing Survey and the $\it Fermi$ Large Area Telescope.
The cross-correlation is a powerful probe of signatures of dark matter 
annihilation, because both cosmic shear and gamma-ray emission 
originate directly from the same DM distribution in the universe,
and it can be used to derive constraints on dark matter annihilation cross-section.
We show that the measured lensing-$\gamma$ correlation is consistent with a null signal. 
Comparing the result to theoretical predictions,
we exclude dark matter annihilation cross sections of $\langle \sigma v \rangle =10^{-24}-10^{-25}
\ {\rm cm}^{3} \ {\rm s}^{-1}$ for a $100$ GeV dark matter. 
If dark matter halos exist down to the mass scale of $10^{-6} M_\odot$, 
we are able to place constraints on the thermal
cross sections $\langle \sigma v \rangle \sim 5 \times 10^{-26} \, {\rm cm}^3 \, {\rm s}^{-1}$ 
for a $10$ GeV dark matter annihilation into $\tau^+ \tau^-$. 
Future gravitational lensing surveys will increase sensitivity to probe 
annihilation cross sections
of $\langle \sigma v \rangle \sim 3 \times 10^{-26} \ {\rm cm}^3 \ {\rm s}^{-1}$ 
even for a 100 GeV dark matter. 
Detailed modeling of the contributions from astrophysical sources to the cross
correlation signal could further improve the constraints by $\sim 40-70$ \%.
\end{abstract}

\maketitle

\section{\label{sec:intro}INTRODUCTION}

The origin of the extragalactic gamma-ray background (EGB) emission is among the 
most interesting problems in astrophysics. The EGB was first detected by the 
$\it OSO$-3 satellite \citep{1972ApJ...177..341K} and subsequently deduced by the 
$\it SAS$-2 satellite \citep{1973ApJ...186L..99F} and the Energetic Gamma-Ray 
Experiment Telescope onboard the Compton Gamma-ray Observatory \citep{Sreekumar:1997un}. 
Most recently,  the Large Area Telescope (LAT) onboard the $\it Fermi$ Gamma-ray 
Space Telescope has derived the most accurate EGB based on new data and improved 
modeling of the  Galactic gamma-ray foreground emission. The $\it Fermi$ LAT 
observation shows a featureless  power-law spectrum for the EGB in the energy range 
$0.1$--$300$ GeV \citep{Abdo:2010nz}.

Multiple astrophysical sources of gamma rays have been proposed as contributors 
to the EGB. Unresolved astrophysical point sources, such as blazars and star-forming 
galaxies (SFG), are guaranteed sources and have been investigated by many groups. 
However, the modeling of the sources' faint end distributions is non-trivial, and estimates 
of the contribution to the EGB from unresolved blazars range from $\sim$15 per cent to 
$\sim$100 per cent \citep[e.g.,][]{Stecker:1996ma, Narumoto:2006qg, Inoue:2008pk}. 
On the other hand, the intrinsic spectral and flux properties of blazars constructed by 
$\it Fermi$ LAT data, as well as the auto-correlation of EGB anisotropies \citep{Ackermann:2012uf}, 
suggest that unresolved blazars can only contribute up to $\sim$20 per cent of 
EGB \citep[e.g.,][]{Collaboration:2010gqa, Ajello:2011zi, Harding:2012gk, Cuoco:2012yf}. 
Similarly, the contribution from SFGs and radio galaxies to the EGB can 
be significant but is subject to large uncertainties \citep{Ackermann:2012vca, Inoue:2011bm}. 
These previous works show that while the EGB intensity can be explained by the superposition 
of multiple astrophysical source classes, there appears to remain large uncertainties 
and thus, at present, an appreciable contribution from unknown or unconfirmed sources of gamma rays
is allowed. 

Among the potential new contributors to the EGB is the emission due to dark matter (DM)
annihilation. The existence of DM is supported with high significance
by a number of astrophysical observations, such as the cosmic microwave background 
(CMB) anisotropies \citep[e.g.,][]{Komatsu:2010fb, Ade:2013zuv} 
and large-scale structure 
\citep[e.g.,][]{Tegmark:2006az, Anderson:2012sa, Kilbinger:2012qz}. 
While the DM  particle properties still remain 
unclear, if DM particles annihilate into standard model particles, as is typically
expected for their production in the early universe, they will produce gamma rays that
contribute to the observed EGB. The gamma-ray emission due to DM annihilation 
is expected to be anisotropic because of the highly non-linear gravitational growth of the DM 
density distribution
\citep[e.g.,][]{Ando:2005xg}. Although astrophysical sources are also expected to 
reside within DM halos, differences in their clustering properties may help 
distinguish DM annihilation signals from astrophysical contributions to the EGB. 

The DM distribution in the Universe can be traced in a number of ways. Among the
most powerful is gravitational lensing, which has the advantage of not requiring any 
assumptions such as the relation between luminosity and mass and/or hydrostatic 
equilibrium.
The small distortions in images of distant objects caused by the large-scale 
matter distribution 
along the line of sight is called cosmic shear.
The DM distribution that generate 
cosmic shear would also be a gamma-ray source.
The cross-correlation between cosmic shear and the EGB would provide competitive 
information of dark matter annihilation \citep{Camera:2012cj}. In ref.~\cite{Camera:2012cj},
the authors also explored how astrophysical sources contribute 
to the cross-correlation signal, and concluded that even without detailed astrophysical 
modeling, the additional information derived by the cross-correlation 
would be helpful for a unified understanding of the EGB.

In this paper, we present the first measurement of the cross-correlation 
between cosmic shear and the EGB using the largest cosmic shear data set currently 
available from the Canada-France-Hawaii Lensing Survey (CFHTLenS) and gamma-ray 
photon data from the $\it Fermi$ LAT telescope. We carefully remove contributions from 
gamma-ray point sources and the Galactic gamma-ray foreground using a likelihood 
modeling based on official {\it Fermi} tools and Galactic diffuse background models. 
We then determine the cross-correlation signal. Finally, by using statistical errors 
derive from the real cosmic shear and gamma-ray data, 
we place novel and competitive constraints 
on the DM annihilation cross section as functions of the DM mass and annihilation channel. 

The paper is organized as follows.
In Section~\ref{sec:DMann}, we summarize the basics of DM, including the 
contribution to the EGB. 
In Section~\ref{sec:data}, we describe the cosmic shear and gamma-ray data used, 
and provide details of the cross-correlation analysis. 
Our benchmark model of the cross-correlation is discussed in Section~\ref{sec:cross}.
In Section~\ref{sec:res}, we show the result of our cross-correlation analysis,
and discuss constraints on the DM annihilation cross section.
Finally, we forecast DM constraints that can be achieved
with upcoming lensing surveys.
Concluding remarks and discussions are given in Section~\ref{sec:con}. 
Throughout, we use the standard cosmological parameters $H_0=100 h \, {\rm km \, s^{-1}}$
with $h=0.7$, $\Omega_{\rm m0}=0.279$, and $\Omega_{\Lambda}=0.721$.

\section{\label{sec:DMann}Dark Matter Annihilation}

The contribution of DM annihilation to the EGB intensity $I_\gamma$ (the number 
of photons per unit energy, area, time, and solid angle) is
\begin{equation}\label{eq:Intensity}
E_\gamma I_\gamma = \frac{c}{4\pi} \int {\rm d}z \frac{P_\gamma (E'_\gamma,z)}{H(z)(1+z)^4} e^{-\tau(E'_\gamma,z)},
\end{equation}
where $E_\gamma$ is the observed gamma-ray energy, $E'_\gamma = (1+z) E_\gamma$ is the energy 
of the gamma ray at redshift $z$, $H(z) = H_0 [\Omega_{\rm m0}(1+z)^3+\Omega_\Lambda]^{1/2}$ is the 
Hubble parameter in a flat Universe, 
and the exponential factor in the integral
takes into account the effect of gamma-ray attenuation during propagation 
owing to pair creation on diffuse extragalactic photons. 
Although the effect of attenuation is only 
important for photon energies larger than $\sim 1$ TeV, and hence is not of great importance for our 
analysis that focuses on lower energy photons, we include it for completeness. For
the gamma-ray optical depth $\tau\left(E'_\gamma, z \right)$, we adopt the model in 
Ref.~\citep{Gilmore:2011ks}.
Finally, $P_\gamma$ is the volume emissivity (the photon energy 
emitted per unit volume, time, and energy range), which is given by
\begin{equation}\label{eq:dmEmissivity}
P_\gamma(E_\gamma,z)=E_\gamma \frac{{\rm d}N_\gamma}{{\rm d}E_\gamma} 
\frac{\langle \sigma v \rangle}{2} 
\left[ \frac{\rho_{\rm dm}({\bd x}|z)}{m_{\rm dm}} \right]^2,
\end{equation}
where ${\rm d}N_\gamma /{\rm d}E_\gamma$ is the gamma-ray spectrum per annihilation, 
$\langle \sigma v \rangle$ is the annihilation cross section times the relative velocity averaged 
with the velocity distribution function, 
$\rho_{\rm dm}({\bd x}|z)$ is the DM mass density distribution
at redshift $z$ as a function of spatial coordinate ${\bd x}$,
and 
$m_{\rm dm}$ is the DM particle mass. 

For the gamma-ray spectrum per annihilation ${\rm d}N_\gamma / {\rm d}E_\gamma$, we adopt two
characteristic spectra corresponding to annihilation with $100$\% branching ratios to 
$b\bar{b}$ and $\tau^+\tau^-$ final states, using the {\tt PPPC4DMID} package 
\citep{Cirelli:2010xx} that is based on {\tt PYTHIA} (v8.135) and {\tt HERWIG} (v6.510) 
event generators. The spectra are dominated by emission from the decay of neutral pions. 
These are {\it primary} gamma-ray emissions, and are distinguished from {\it secondary} 
emission 
that results from interactions of the annihilation products with the environment. 
An example of the latter is when DM annihilation produces a positron, which, in turn, 
finds an electron in the galactic halo and annihilates to produce gamma rays. Also, the 
gamma-ray emission can be noticeably softened by the bremsstrahlung emission
from leptonic final states \citep{Cirelli:2013mqa}. We do not include secondary emission 
in this study because their effect depends strongly on the astrophysical environment
and furthermore since they are only critical for annihilation in regions of high baryon density, 
e.g., the planes of galaxies. Additional contributions can arise from three-body final states such as
internal bremsstrahlung \citep{Bergstrom:2005ss}. This would introduce a sharp feature 
near the DM mass and systematically harden the gamma-ray emission. The sharp feature 
may enhance the correlation signal and provide a useful diagnostic for DM annihilation;
it has been discussed in the context of anisotropies \citep{Zhang:2004tj,Campbell:2013rua}.
However,  we do not include this because it can only be included in the framework of 
a precise DM model \citep[e.g.,][]{Cirelli:2010xx}. 

Since the DM annihilation rate scales with the DM density squared, highly 
over-dense regions such as DM halos dominate 
the volume emissivity. It is instructive to express the DM density $\rho_{\rm dm}$ as an 
overdensity $\delta(z) = \rho_{\rm dm} / \bar{\rho}_{\rm dm}(z)$ over the mean DM density 
$\bar{\rho}_{\rm dm}(z)=\Omega_{\rm dm} \rho_{\rm crit}(1+z)^3$, where $\rho_{\rm crit}$ is the critical 
density. The ensemble average of the overdensity squared, 
$\langle \delta^2(z)\rangle=\langle \rho_{\rm dm}^2(z)\rangle/ \bar{\rho}_{\rm dm}^2(z)$, is called the 
intensity multiplier (or the clumping factor), and characterizes the 
enhancement in the DM annihilation rate due to dense DM halos. 
It is obtained by integrating over the DM halo mass function $n(M,z)$,
\begin{equation}\label{eq:clumping}
\langle \delta^2(z)\rangle = \frac{1}{\bar{\rho}^2_{\rm dm}(z)} \int^\infty_{M_{\rm min}} {\rm d}M 
\,n(M,z) \int {\rm d}V \rho_{\rm dm}^2(r|M,z),
\end{equation}
where
$\rho_{\rm dm}(r|M,z)$ describes the density profile as a function
of radius $r$ for a DM halo with mass $M$ at redshift $z$, 
and $M_{\rm min}$ is the smallest DM halo mass. 

Estimates of the flux multiplier depend on the value of $M_{\rm min}$, the halo mass 
function, the DM density profile, and how the DM profile depend on halo mass and 
evolve in redshift. Among these, the value of $M_{\rm min}$ has the largest impact. The 
smallest DM halo mass ought to be determined from the DM particle properties, being the
Jeans mass of dark matter particles. For supersymmetric neutralinos and $\sim$ MeV
DM, this is some $10^{-6} M_\odot $ \citep{Rasera:2005sa}, while other DM particles 
have $M_{\rm min}$ that vary by orders of magnitudes 
\citep{Hofmann:2001bi,Loeb:2005pm,Bringmann:2009vf}. However, complications 
arise because not all DM halos survive the process of mergers and tidal interactions 
during structure formation. In particular, much of the smallest DM halos may be 
absorbed into larger halos and their central densities disrupted before they appreciably 
contribute to the EGB \citep[e.g.,][]{Goerdt:2006hp,Berezinsky:2007qu}.
The DM Jeans mass is therefore
simply a lower limit. Furthermore, for secondary gamma-ray emission, the relevant 
minimum mass is set by the Jeans mass of the baryons, which is on the order of 
$\sim 10^6 M_\odot$ \citep[e.g.,][]{Rasera:2005sa}.  In Section \ref{subsec:halomodel}, 
we discuss the details of the calculation of the flux multiplier in the halo model approach, 
and also discuss the effects of substructures residing within halos. 

Using the flux multiplier, the contribution to the EGB is
\begin{equation}\label{eq:clumping}
I_\gamma = \frac{\langle \sigma v \rangle}{8 \pi} \int c \, {\rm d}z 
\frac{dN_\gamma}{dE_{\gamma}} 
\Bigg|_{E^{\prime}_{\gamma}}
\frac{e^{-\tau(E'_\gamma,z)}}{H(z) (1+z)^3} \left( \frac{\bar{\rho}_{\rm dm}(z)}{m_{\rm dm}} \right)^2 
\langle \delta^2(z)\rangle,
\end{equation}
where the particle properties of DM -- $m_{\rm dm}$, $\langle \sigma v \rangle$, and 
${\rm d}N_\gamma / {\rm d}E_\gamma$ -- are conveniently decoupled from the physics determining 
its spatial distribution, $\langle \delta^2(z)\rangle$. 

\section{\label{sec:data}DATA}

\subsection{Cosmic shear data}
We use the cosmic shear data from the Canada-France-Hawaii Telescope Lensing Survey
 \citep[CFHTLenS;][]{Heymans:2012gg}. CFHTLenS is a 154 square deg multi-color 
 optical survey in five optical bands $u^{*}, g^{\prime}, r^{\prime}, i^{\prime}, z^{\prime}$. 
 CFHTLenS is optimized for weak lensing analysis with a full multi-color depth of 
 $i^{\prime}_{AB} = 24.7$ with optimal sub-arcsec seeing conditions. The survey consists of 
 four separated fields called W1, W2, W3, and W4, with an area of $\sim$ 72, 30, 50, and
  25 square degs, respectively.

The CFHTLenS survey analysis mainly consists of the following three processes:  
photometric redshift measurement \citep{Hildebrandt:2011hb}, weak 
lensing data processing with THELI \citep{Erben:2012zw}, and shear measurement with 
$lens$fit \citep{Miller:2012am}.
A detailed systematic error study of the shear measurements in combination with the 
photometric redshifts is presented in Ref.~\citep{Heymans:2012gg}
and additional error analyses of the photometric redshift measurements are presented 
in Ref.~\citep{Benjamin:2012qp}.

The ellipticities of the source galaxies in the data have been calculated using the $lens$fit 
algorithm. $lens$fit performs a Bayesian model fitting to the imaging data by varying a galaxy's 
ellipticity and size, and by marginalizing over the centroid position. It adopts a 
forward convolution process that convolves the galaxy model with the point spread
function (PSF) to estimate the posterior probability of the model given the data. For each 
galaxy, the ellipticity $\epsilon$ is estimated as the mean likelihood of the model posterior 
probability after marginalizing over galaxy size, centroid position, and bulge fraction. An 
inverse variance weight $w$ is given by the variance of the ellipticity likelihood surface 
and the variance of the ellipticity distribution of the galaxy population.
The $lens$fit algorithm has been tested with image simulations in detail.
The observed ellipticities ${\bd \epsilon}^{\rm obs}$ with any shape measurement method
are calibrated in practice as
\beqa
{\bd \epsilon}^{\rm obs}=(1+m){\bd \epsilon}^{\rm true}+{\bd c},
\eeqa
where $m$ is a multiplicative bias and ${\bd c}$ is an additive bias.
In the case of $lens$fit, ${\bd c}$ is consistent with zero for a large set of simulated images
but $m$ cannot be negligible and it depends on both galaxy signal-to-noise ratio and size.
On a weight average, this multiplicative bias corresponds to a 6 \% correction.
In terms of statistical quantities such as two point correlation function,
this bias is easily corrected by multiplying an overall factor
(see Ref.~\citep{Miller:2012am} 
 for further details).

The PSF in optical imaging surveys is one of the major systematics of 
galaxy shape measurement.
The optical PSF originates from diffraction, the atmospheric turbulence,
optical aberration, the misalignment of CCD chips on a focal plane,
and pixelization effects. 
Anisotropy of the PSF causes a coherent deformation of images
that might mimic the tangential shear pattern due to large scale structure 
in the universe.
Often in cosmic shear measurement, 
systematic effects are tested through statistical analyses of 
the $45^\circ$ rotated component of galaxy ellipticities.
This is because the $45^\circ$ rotated component of cosmic shear should vanish statistically.
In Section \ref{sec:res}, we perform statistical analysis 
by using the $45^\circ$ rotated component of galaxy ellipticities
and we quantify systematics, if any, of the lensing data set.
 
The photometric redshifts $z_{p}$ are estimated by the {\tt BPZ} code \cite[][Bayesian
 Photometric Redshift Estimation]{Benitez:1998br}. 
 The true redshift distribution is well described by the sum of the probability distribution 
 functions (PDFs) estimated from {\tt BPZ} \citep{Benjamin:2012qp}. 
 The galaxy-galaxy-lensing redshift scaling analysis 
 confirms that contamination is not significant for galaxies selected 
 at $0.2 < z_{p} < 1.3$ \citep{Heymans:2012gg}. 
 In this redshift range, the weighted median redshift is $\sim0.7$ and 
 the effective weighted number density $n_{\rm eff}$ is 11 per square arcmin. 
 We have used the source galaxies with $0.2 < z_{p} < 1.3$ to measure the cross-correlation of 
 cosmic shear and EGB presented in Section~\ref{sec:cross}. 
 We use a total of 2570270, 679070, 1649718, 
 and 770356 galaxies in the W1, W2, W3, and W4 fields, respectively, for our cross-correlation study. 

\subsection{Extragalactic gamma-ray background data}
We use ${\it Fermi}$-LAT Pass 7 Reprocessed gamma-ray photon data taken from August 2008 
to January 2014. For each CFHTLenS patch, we download photons within a circle of radius 
$10^\circ$ around the center of each region and work with a $14^\circ \times 14^\circ$ 
square region of interest (ROI). We use the Fermi Tools version {\tt v9r32p5} to analyze the 
data\footnote{http://fermi.gsfc.nasa.gov/ssc/data/analysis/}. Using the {\it gtmktime} tool, 
we remove data taken during non-survey modes and 
when the satellite rocking angle exceeds $52^\circ$ with respect to the zenith ({\tt DATA\_QUAL=1},
{\tt LAT\_CONFIG=1}, and {\tt ABS(ROCK\_ANGLE)<52}). This standard 
procedure removes epochs with potentially significant contamination by the gamma-ray bright
Earth limb. Unless otherwise stated, we work with only {\tt ULTRACLEAN}-class 
photons, which are events that pass the most stringent quality cuts, and we use photons between
$1$--$500$ GeV in energy. In Section \ref{subset:analysis}, we discuss using {\tt SOURCE}-class 
photons. We use the {\it gtbin} tool to bin the photons in a stereographic projection 
into pixels of $0.2^\circ \times 0.2^\circ$ and into $30$ equal logarithmically-spaced energy bins. 
These binning sizes are taken from the official recommended values that are chosen to ensure 
reasonable analysis outcomes, namely, to ensure that rapid variations of the effective area with 
energy is taken into account (e.g., as discussed in the binned likelihood tutorial of the Fermi
Analysis Threads). With the data selection cuts in place, we use the {\it gtltcube} 
tool to generate integrated live times and the {\it gtexpcube2} tool to generate the integrated 
exposure maps. Throughout, we work with the {\tt P7REP\_ULTRACLEAN\_V15} instrument 
response function (IRF), unless otherwise stated.  

In order to obtain the extragalactic diffuse photons, for each ROI we subtract the best fit Galactic
foreground emission model from the raw data. We then mask out point sources using a mask of
$2^\circ$ radius around each point source. The mask size corresponds to a generous estimate 
of the PSF of the ${\it Fermi}$-LAT detector, which decreases with energy: the 
$68$\% containment angle is $\sim 0.9$ deg at $1$ GeV and $\sim 0.26$ deg at $10$ 
GeV, both for combined front and back conversion tracks. Since most point sources have steep
spectra and hence dominated by low-energy photons, our adopted mask is chosen to be sufficiently
larger than the containment angle at our lower energy limit of $1$ GeV. 
We discuss the potential of smaller mask sizes in Section~\ref{sec:con}. 

\begin{figure}[!t]
\begin{center}
       \includegraphics[clip, width=0.45\columnwidth]{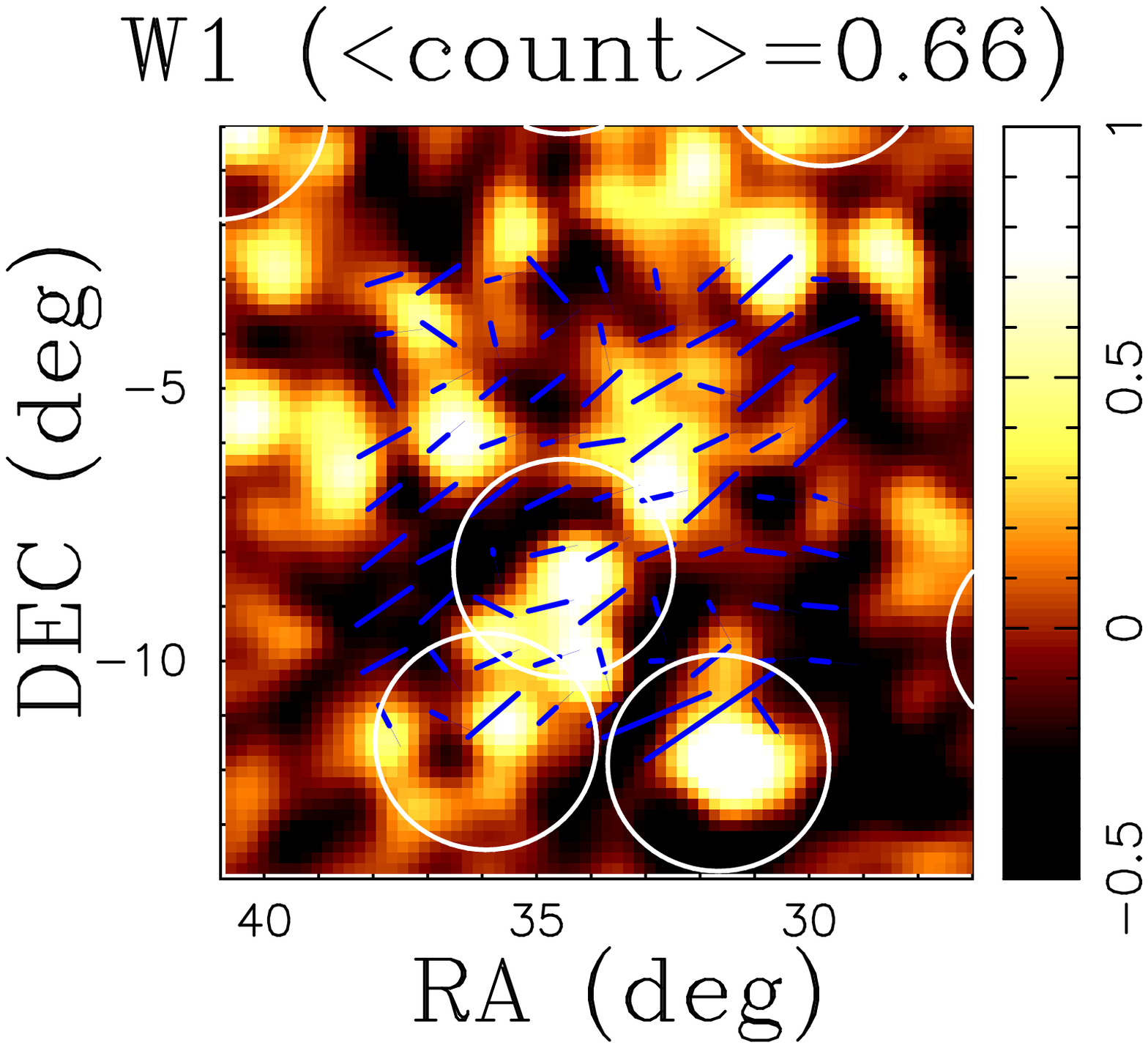}
       \includegraphics[clip, width=0.45\columnwidth]{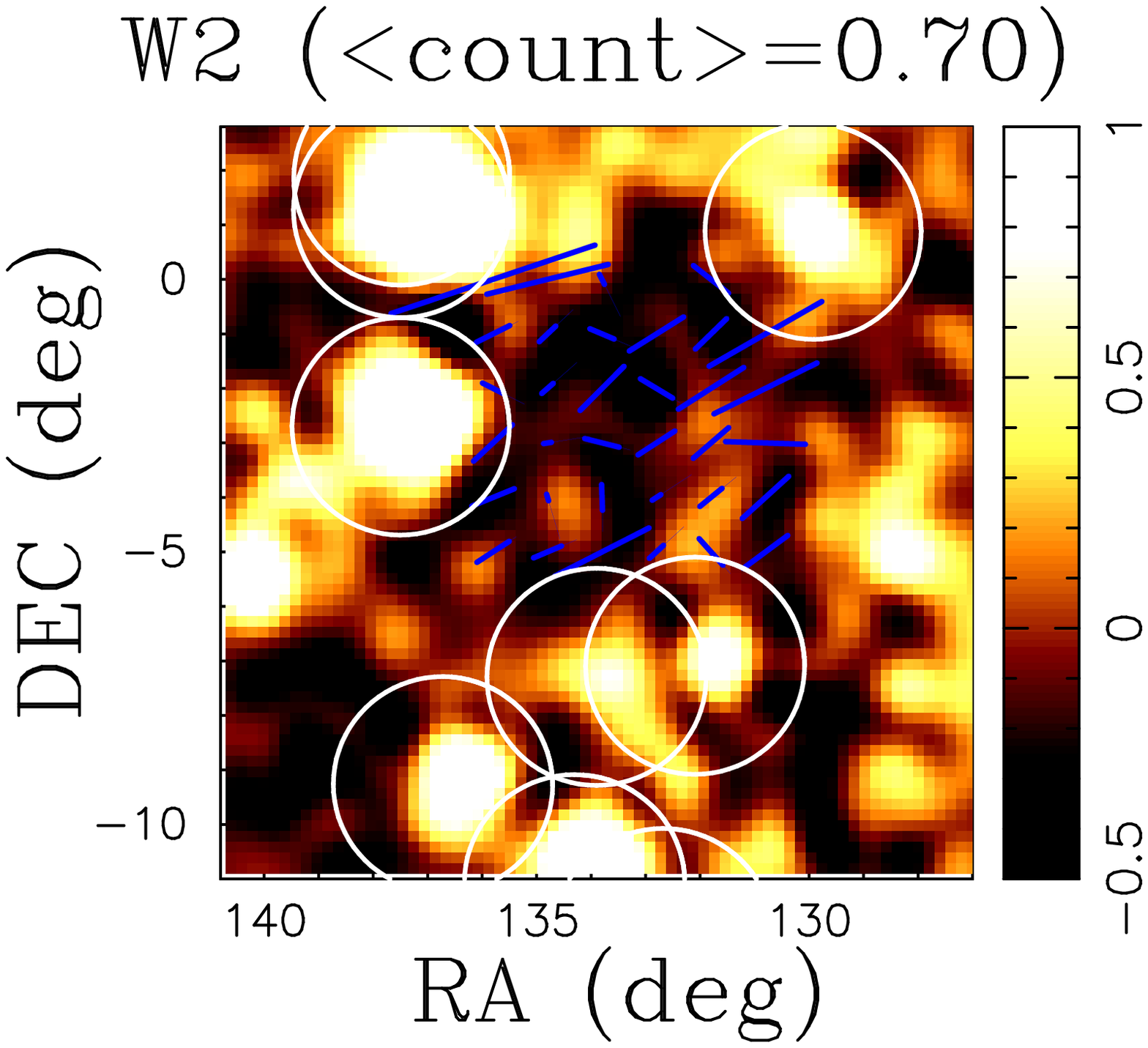}
       \includegraphics[clip, width=0.45\columnwidth]{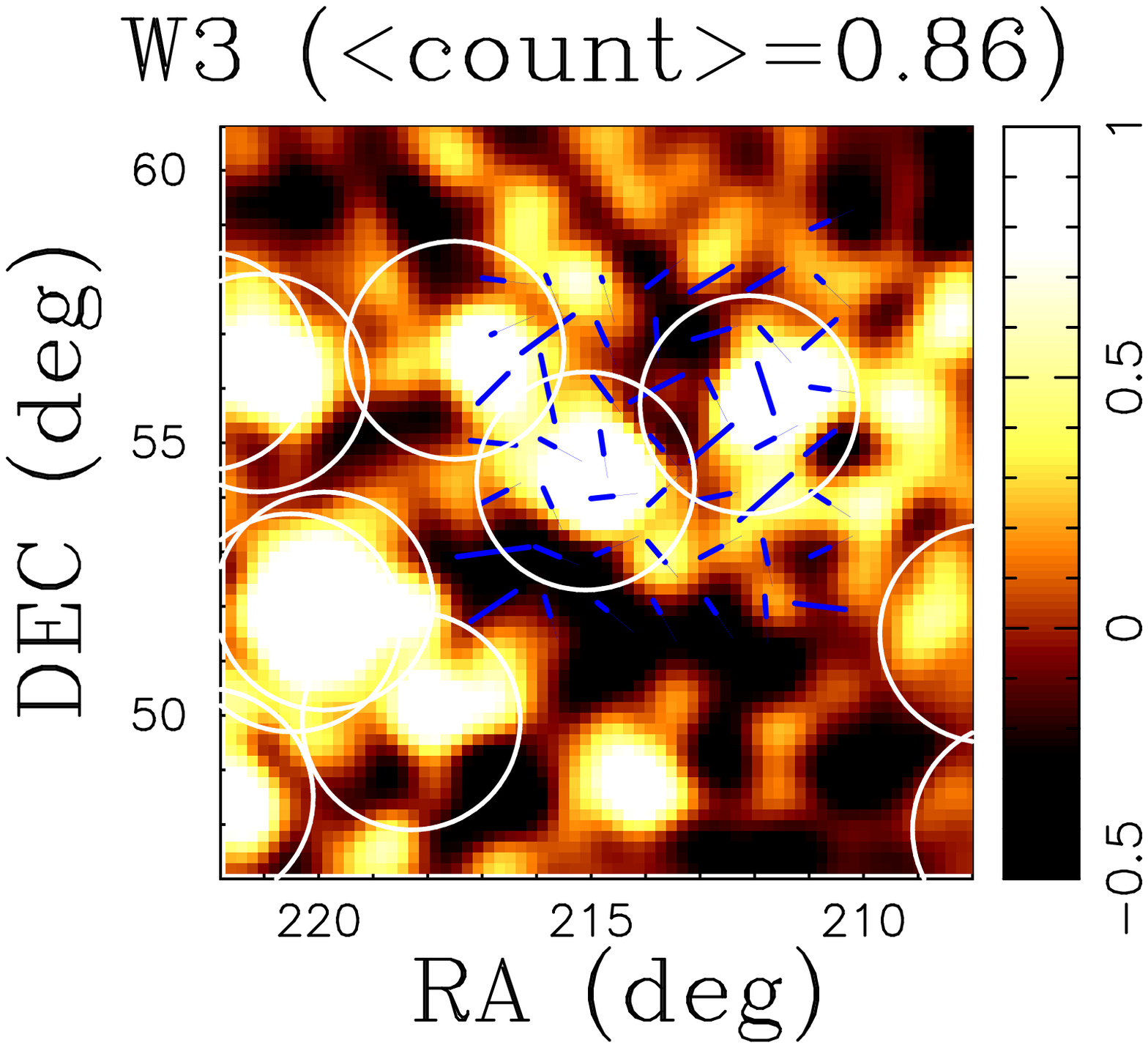}
       \includegraphics[clip, width=0.45\columnwidth]{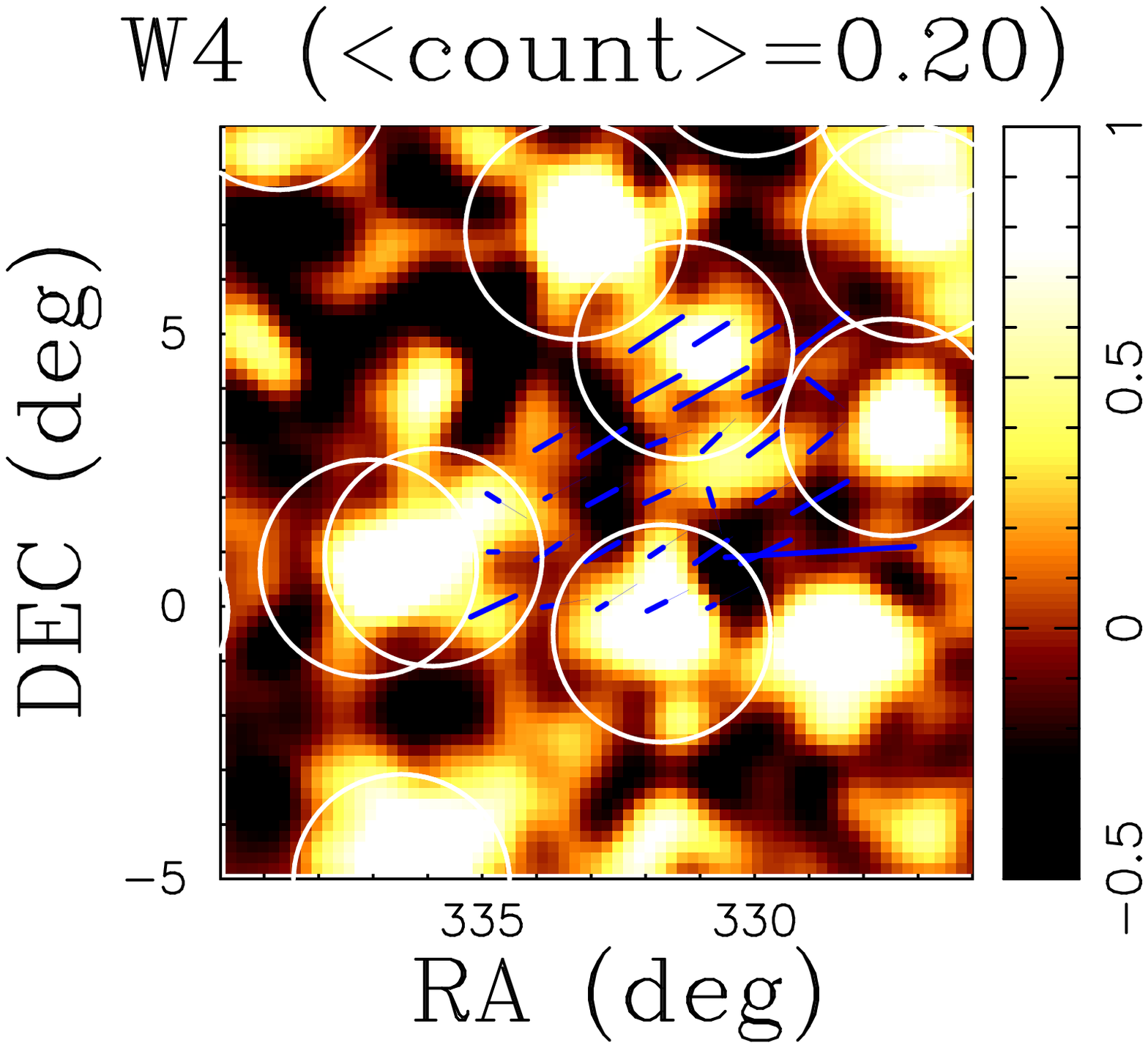}       
    \caption{
    \label{fig:W1-4}
    Residual maps in the CFHTLenS W1, W2, W3, 
    and W4 fields, where residual is defined 
    as the fluctuation in the EGB photon count map 
    from its mean value. 
    In each panel, the color-scale bar shows both the positive and negative 
    difference between the EGB count map and the mean of each field 
    indicated above the panels: 0.66, 0.70, 0.86, and 0.20 in W1, W2, 
    W3 and W4 fields, respectively. Overlaid by thick lines are the average 
    ellipticities of source galaxies over 1 deg$^2$ with arbitrary scaling. The 
    circles represent the point-source masked regions. For visualization
    purposes, a Gaussian smoothing is performed on the map with a width of 0.6 deg.
    }
    \end{center}
\end{figure}

The best fit Galactic diffuse emission model is estimated separately for each ROI, by including all 
the point sources in the ROI in the 2FGL catalog, together with the recommended Galactic diffuse 
emission model ({\tt gll\_iem\_v05}) and the recommended isotropic emission model 
({\tt iso\_clean\_v05}). 
We have checked that our four ROIs are sufficiently far from the large-scale diffuse gamma-ray sources
such as the Fermi bubbles \citep{Su:2010qj} which would otherwise complicate fitting.
The CFHTLenS patches each have 9, 11, 11, and 12 point sources, 
respectively. We use the {\it gtlike} tool to perform a binned likelihood analysis, varying all point 
source spectra as well as the diffuse emission normalizations. We then use the {\it gtmodel} tool 
to generate photon counts maps based on the best fit Galactic diffuse model and exposure maps. 
Finally, we subtract these from the raw counts maps. We checked that the procedure yields a flux 
spectrum for the EGB, estimated as the raw counts minus a model without the isotropic component, 
divided by the exposure map, that is very similar to the $-2.41$ power-law spectrum of the EGB 
reported in Ref.~\citep{Abdo:2010nz}. 
In Figure \ref{fig:W1-4}, we show how the residuals of the raw counts minus the Galactic diffuse model,
demonstrate structureless spatial maps in all four CFHTLenS fields.

\section{\label{sec:cross}CROSS-CORRELATION OF COSMIC SHEAR AND EGB}

\subsection{\label{subset:analysis}Analysis}

In order to calculate the cross-correlation of cosmic shear and EGB, 
we use the following estimator: 
\beqa
\xi_{\delta n-\gamma_{t}}(\theta)
=\frac{\displaystyle \sum^{N_{\rm pixel}}_{i}\sum^{N_{\rm gal}}_{j}(n^{\rm obs}({\bd \phi}_{i})-n^{\rm gm}({\bd \phi}_{i}))w_{j}\epsilon_{t}({\bd \phi}_{j}|{\bd \phi}_{i})\Delta_{\theta}({\bd \phi}_{i}-{\bd \phi}_{j})}{(1+K(\theta))\displaystyle \sum^{N_{\rm pixel}}_{i}\sum^{N_{\rm gal}}_{j}w_{j}\Delta_{\theta}({\bd \phi}_{i}-{\bd \phi}_{j})},
\label{eq:CCest}
\eeqa
where $N_{\rm pix}$ is the number of pixels in the gamma-ray counts map,
$N_{\rm gal}$ is the number of galaxies,
$n^{\rm obs}({\bd \phi}_{i})$ is the observed number of photons in pixel $i$ in the gamma-ray 
counts map, $n^{\rm gm}({\bd \phi}_{i})$ is the contribution from the Galactic emission model estimated 
using the ${\it Fermi}$-LAT diffuse template and detector modeling, $w_{j}$ is the weight related to the shape measurement, 
and $\epsilon_{t}({\bd \phi}_{j}|{\bd \phi}_{i})$ is the tangential component of the $j$-th galaxy's 
ellipticity with respect to the $i$-th pixel of the gamma-ray counts map, defined by
\beqa
\epsilon_{t}({\bd \phi}_{j}|{\bd \phi}_{i}) = -\epsilon_{1}({\bd \phi}_{j})\cos(2\alpha_{ij})-\epsilon_{2}({\bd \phi}_{j})\sin(2\alpha_{ij}),
\eeqa
where $\alpha_{ij}$ is defined as the angle measured from the right ascension direction to a line 
connecting the $i$-th pixel and the $j$-th galaxy. 
We define the function $\Delta_{\theta}({\bd \phi}) = 1$ 
for $\theta-\Delta \theta/2 \le \phi \le \theta+\Delta \theta/2$ and zero otherwise.
The overall factor $1+K(\theta)$ in Eq.~(\ref{eq:CCest}) 
is used to correct for the multiplicative shear bias $m$ in 
the shape measurement with $lens$fit \citep{Miller:2012am}, which is given by 
\beqa
1+K(\theta)=
\frac{\displaystyle \sum^{N_{\rm pixel}}_{i}\sum^{N_{\rm gal}}_{j}
w_{j}(1+m({\bd \phi}_{j}))\Delta_{\theta}({\bd \phi}_{i}-{\bd \phi}_{j})}
{\displaystyle \sum^{N_{\rm pixel}}_{i}\sum^{N_{\rm gal}}_{j}w_{j}\Delta_{\theta}({\bd \phi}_{i}-{\bd \phi}_{j})}.
\eeqa
We have checked that our estimator is consistent with a zero signal
when applied to randomized shear catalogues 
and the observed photon count map. We have also tested
a combination of random photon count map with the observed shear catalogue.

For binning in angular separation $\theta$, we set the innermost separation bin to 1 arcmin and use 
10 bins logarithmically spaced in $\Delta \log_{10}\theta = 0.2$. 
In calculating Eq.~(\ref{eq:CCest}), we do not perform pixelization 
in the galaxy catalogue.
We simply consider the center of each 
pixel in the gamma-ray map as the angular position of the gamma-ray photons 
to perform the summation in Eq.~(\ref{eq:CCest}).
To be precise, this induces an artificial smoothing over smaller scales than the 
pixel size in our gamma-ray map, 
i.e., 0.2 deg.
However, we do not expect to detect physically important correlations over such small angular 
scales due to blurring by 
the PSF of the ${\it Fermi}$-LAT detector, as we show in Section \ref{subsec:halomodel}.
In the present paper,
we take the PSF smearing into account in theoretical models (see Figure 4).
Note that the pixelization effect in the gamma-ray map is included
in the covariance of our estimator.
The pixelization effect is found to be unimportant
in detection of the cross-correlation signals at large angular separations.

\begin{figure}[!t]
\begin{center}
       \includegraphics[clip, width=0.6\columnwidth]{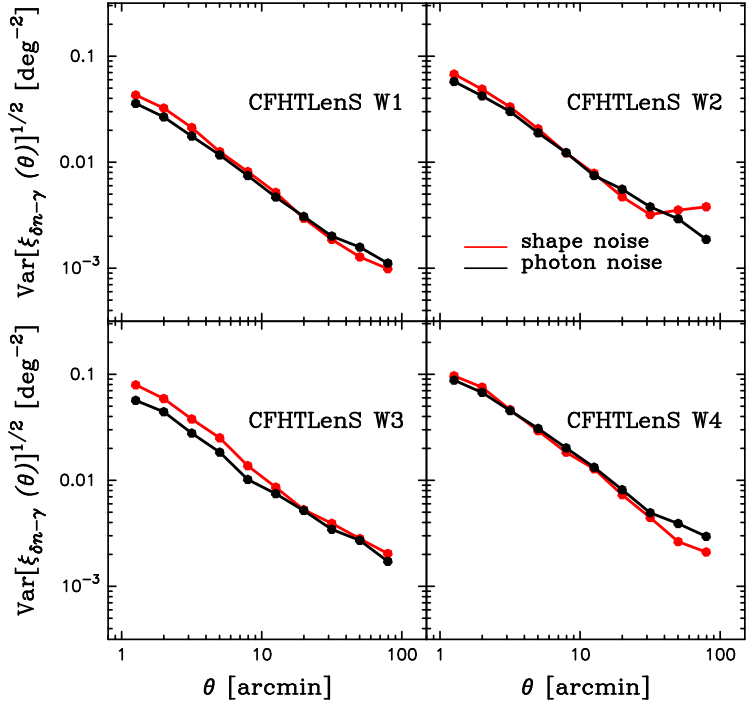}
       
    \caption{
    \label{fig:var_xi}
   The variance of cross-correlation signals
   estimated from a set of randomized realizations and the observed map.    
   The red line in each panel represents the statistical error associated with
   the shape measurement.
   The black line shows 
   the statistical error associated with
   the Poisson error from the finite number of gamma-ray counts.
  } 
    \end{center}
\end{figure}

The statistical properties of our estimator Eq.~(\ref{eq:CCest})
are summarized in Appendix \ref{sec:appendixA}.
There, we present the exact formulation of the covariance of our estimator and
derive two dominant contributions;
they arise from the intrinsic shape variance of galaxies, called shape noise,
and the finite number of photon counts per pixel in the gamma-ray maps, called photon noise.
We use randomized shear catalogues
in order to estimate the statistical errors associated with the shape noise. 
To this end, we generate 500 
randomized shear catalogues by rotating the direction of each galaxy ellipticity 
but with fixed 
amplitude \citep{Shirasaki:2013zpa}. 
We then estimate the covariance matrix $C_{ij}$ of the estimator Eq.~(\ref{eq:CCest}) by
\beqa
C_{ij} = \frac{1}{N_{\rm re}-1}\sum_{r}(\xi_{\delta n-\gamma_{t}}^{r}(\theta_{i})-{\bar \xi}_{\delta n-\gamma_{t}}(\theta_{i}))(\xi_{\delta n-\gamma_{t}}^{r}(\theta_{j})-{\bar \xi}_{\delta n-\gamma_{t}}(\theta_{j})),\label{eq:cov}
\eeqa
where $\xi_{\delta n-\gamma_{t}}^{r}(\theta_{i})$ 
is the estimator for the $i$-th angular bin obtained 
from the $r$-th realization, 
and $N_{\rm re} = 500$ is the number of randomized catalogues. 
The ensemble average of the $i$-th angular bin over 500 realizations, 
${\bar \xi}_{\delta n-\gamma_{t}}(\theta_{i})$, is 
simply given by
\beqa
{\bar \xi}_{\delta n-\gamma_{t}}(\theta_{i})
=\frac{1}{N_{\rm re}}\sum_{r}\xi_{\delta n-\gamma_{t}}^{r}(\theta_{i}).
\eeqa
To simulate the photon count noise, 
we generate 500 randomized count maps 
assuming the photon counts in each pixel follows a Poisson distribution 
with a mean of $n^{\rm obs}({\bd \phi})$.
We repeat the cross-correlation analysis with the 500 count maps 
and the observed galaxy shear catalogue.
We then estimate the statistical error related to the photon noise 
in the same manner shown in Eq.~(\ref{eq:cov}).
In total, we estimate the statistical error associated 
with the shape measurement and the photon noise
by summing these two contributions.
Figure \ref{fig:var_xi} shows the variance of the cross-correlation signal 
estimated from the two sets of randomized realizations as described above.
In each panel, the red line shows the contribution from the shape noise
and the black line shows the variance due to the photon noise.
Overall, the shape noise and the photon noise contribute to the statistical error of our estimator
at similar levels.

The cross-correlation estimator is also dependent on the model for the 
foreground astrophysical diffuse emission of our own Galaxy. We therefore investigate 
alternate LAT diffuse models provided by the ${\it Fermi}$ collaboration to assess differences
in the estimated EGB photons. First we work with  ${\it Fermi}$ LAT Pass 7 reprocessed 
{\tt SOURCE}-class photons, which is made with a weaker set of cuts to remove cosmic-ray induced 
backgrounds, and analyze them adopting the appropriate diffuse model and IRF. Second, 
we work with the ${\it Fermi}$ LAT Pass 7 photon pipeline rather than Pass 7 reprocessed 
photons with respectively the appropriate diffuse emission model
({\tt gal\_2yearp7v6\_v0} and {\tt iso\_p7v6clean}) and IRF. In both cases, we first find 
the best fit diffuse model normalizations, subtract the best fit Galactic diffuse maps from 
the raw data, and then mask the point sources, to obtain finally the EGB photons. We have explicitly
checked that the 
different Galactic diffuse models do not significantly affect our cross-correlation analyses. 
We discuss this issue later in Section \ref{sec:con}. 

It may be necessary to consider another important contribution to the covariance, i.e., the sampling variance. 
To estimate the sampling variance, one could use the halo model approach 
 \ref{subsec:halomodel}, but it is uncertain how the astrophysical sources are included
in the model. Because we expect the sampling variance to be less important compared to the
uncertainty of the halo model $\it itself$,
we simply ignore the sampling 
variance but include the model uncertainty as presented in \ref{subsec:halomodel} when 
deriving the constraints on DM annihilation.

\subsection{\label{subsec:halomodel}Theoretical model}
In this section, we summarize our benchmark model for the cross-correlation signal between cosmic
shear and the EGB. The theoretical framework for the angular power spectrum analysis of the 
EGB has been developed in Refs.~\citep{Ando:2005xg, Ando:2006mt, Ando:2013ff, Camera:2012cj}. 
We calculate the cross-correlation of cosmic shear and the EGB as follows.
  
In general, the number of EGB photons along the line of sight ${\bd \theta}$ can be expressed by
\beqa
\delta n({\bd \theta}) = \int {\rm d}\chi \ g(\chi, {\bd \theta}) W_{g}(\chi), \label{eq:cmap}
\eeqa
where $\chi$ is the comoving distance, $g$ is the relevant field for gamma-ray sources,	
and $W_{g}$ is the window function.
In the case of gamma-ray emission from DM annihilation, 
the relevant field is the overdensity squared $\delta^2$, and 
the window function is given by
\beqa
W_{g}(\chi) = \int_{E_{\gamma, \rm min}}^{E_{\gamma, \rm max}} {\rm d}E_{\gamma}
 \ \frac{\langle \sigma v \rangle}{8\pi}
\left(\frac{{\bar \rho}_{\rm dm, 0}}{m_{\rm dm}}\right)^2
\left[1+z(\chi)\right]^3 \frac{{\rm d}N_{\gamma}}{{\rm d}E_{\gamma}}
\Bigg|_{E^{\prime}_{\gamma}}
\exp \left[ -\tau \left(E^{\prime}_{\gamma}, \chi \right)\right] \eta(E_{\gamma}), 
\label{eq:Wg_dm}
\eeqa
where ${\bar \rho}_{\rm dm,0}$ is the mean density of DM at present, $E^{\prime}_{\gamma}=(1+z(\chi))E_{\gamma}$ 
and $E_{\gamma}$ are the
energy of the gamma ray when it is emitted at $\chi$ and when it is observed, respectively, 
and $\eta(E_{\gamma})$ is the exposure which is the integral of effective area over time taking into
account the orbits of {\it Fermi} and data cuts. We use a standard model of $\tau$ 
\citep{Gilmore:2011ks}, and we estimate $\eta(E_{\gamma})$ by averaging the exposure maps over 
the ROI in each of the CFHTLenS patches.

We next consider gravitational lensing by large-scale structure.
When one denotes the observed position of a source object as $\bd{\theta}$ 
and the true position as $\bd{\beta}$,
one can characterize the distortion of image of a source object by the 
following 2D matrix:
\beqa
A_{ij} = \frac{\partial \beta^{i}}{\partial \theta^{j}}
           \equiv \left(
\begin{array}{cc}
1-\kappa -\gamma_{1} & -\gamma_{2}  \\
-\gamma_{2} & 1-\kappa+\gamma_{1} \\
\end{array}
\right), \label{distortion_tensor}
\eeqa
where $\kappa$ is convergence and $\gamma$ is shear.
In the weak lensing regime (i.e., $\kappa, \gamma \ll 1$), 
each component of $A_{ij}$ can be related to
the second derivative of the gravitational potential $\Phi$ as
\beqa
A_{ij} &=& \delta_{ij} - \Phi_{ij}, \label{eq:Aij} \\
\Phi_{ij}  &=&\frac{2}{c^2}\int _{0}^{\chi}{\rm d}\chi^{\prime} f(\chi,\chi^{\prime}) 
\frac{\partial^2}{\partial x_{i}\partial x_{j}}\Phi[r(\chi^{\prime})\bd{\theta},\chi^{\prime}], \label{eq:shear_ten}\\	
f(\chi,\chi^{\prime}) &=& \frac{r(\chi-\chi^{\prime})r(\chi^{\prime})}{r(\chi)}
\eeqa
where $r(\chi)$ is angular diameter distance, 
and $x_{i}=r\theta_{i}$ represents physical distance \citep[]{Bartelmann:1999yn,Munshi:2006fn}.
By using the Poisson equation, one can relate the 
convergence field to the matter overdensity field $\delta$
\citep[]{Bartelmann:1999yn,Munshi:2006fn}.
Weak lensing convergence field is then given by
\beqa
\kappa(\bd{\theta},\chi)= \frac{3}{2}\left(\frac{H_{0}}{c}\right)^2 \Omega_{\rm m0}
\int _{0}^{\chi}{\rm d}\chi^{\prime} f(\chi,\chi^{\prime}) 
\frac{\delta[r(\chi^{\prime})\bd{\theta},\chi^{\prime}]}{a(\chi^{\prime})}. \label{eq:kappa_delta}
\eeqa
Because source galaxies are distributed over a range of redshift,
we denote the source distribution by $p(\chi)$.
In this case, convergence field on the ${\bd \theta}$ coordinate is expressed as
\beqa
\kappa(\bd{\theta})=\int {\rm d}\chi \,W_{\kappa}(\chi) \delta({\bd \theta}, \chi), \label{eq:kappa_delta_p}
\eeqa
where window function for $\kappa$ is given by
\beqa
W_{\kappa}(\chi) = \frac{3}{2}\left(\frac{H_{0}}{c}\right)^2 \Omega_{\rm m0}(1+z(\chi))
\int_{\chi}^{\infty} {\rm d}\chi^{\prime} \ p(\chi^{\prime})f(\chi^{\prime}, \chi).
\eeqa
In this paper, for $p(\chi)$, 
we use the sum of the posterior probability distribution function of photometric redshift 
\citep{Kilbinger:2012qz,Shirasaki:2013zpa}.

Using Eqs.~(\ref{eq:cmap}) and (\ref{eq:kappa_delta_p})
with Limber approximation \citep{Limber:1954zz,Kaiser:1991qi},
we obtain the angular cross power spectrum of $\delta n$ and $\kappa$ as
\beqa
P_{\delta n-\kappa}(\ell) 
=\int \frac{{\rm d}\chi}{\chi^2} W_{g}(\chi)W_{\kappa}(\chi) P_{\delta-\delta^2}(\ell/\chi, z(\chi)).\label{eq:crosspk}
\eeqa
The direct observable in the present study 
is the cross-correlation function in real space,
which is calculated as
\beqa
\xi_{\delta n-\gamma_{t}}(\theta) = \int \frac{{\rm d}\ell \ell}{2\pi} P_{\delta n-\kappa}(\ell) J_{2}(\ell \theta),
\eeqa
where $J_{2}(x)$ represents the second-order Bessel function
\citep{dePutter:2010jz, Oguri:2010vi}.

The integrand $P_{\delta-\delta^2}(k, z)$ in Eq.~(\ref{eq:crosspk}) is calculated
by following the so-called halo model approach \citep{Cooray:2002dia}. The halo model is a useful approach for incorporating the non-linear growth of the overdensity $\delta$ that determines 
the anisotropy of the EGB. With the halo model approach, 
$P_{\delta-\delta^2}(k, z)$ can be expressed as a sum of two terms 
called the one-halo term and the two-halo term. The former represents the two-point 
correlation within a given DM halo, and the latter corresponds to the correlation due to 
clustering of DM haloes. These two terms can be written as, respectively,
\beqa
P^{1h}_{\delta-\delta^2}(k, z)&=&\left(\frac{1}{{\bar \rho}_{m}}\right)^3
\int_{M_{\rm min}} {\rm d}M \ n(M, z) M \ u(k|M,z) \nonumber \\ 
&&\ \ \ \ \ \ \ \ \ \ \ \ \ \ \ \ \ \ \times (1+b_{sh}(M)) v(k|M,z) \int {\rm d}V \rho^2_{h}(r|M, z), 
\label {eq:1h}\\  
P^{2h}_ {\delta-\delta^2}(k, z)&=&P^{\rm lin}(k, z) \left(\frac{1}{{\bar \rho}_{m}}\right)^3
\left[\int_{M_{\rm min}} {\rm d}M \ n(M,z) b_{h}(M, z) M \ u(k|M,z)\right] \nonumber \\
&&\times \left[\int_{M_{\rm min}} {\rm d}M \ n(M,z) b_{h}(M, z) (1+b_{sh}(M)) v(k|M,z)\int {\rm d}V \rho^2_{h}(r|M, z) \right],
\label {eq:2h}
\eeqa
where $n(M, z)$ is the halo mass function, and
$b_{h}(M, z)$ is the linear halo bias \citep{Sheth:1999mn,Sheth:1999su}.
We adopt the Navarro-Frenk-White (NFW) DM density profile \citep[]{Navarro:1996gj},
\beqa
\rho_{h}(r|M, z) = \frac{\rho_{s}}{\left(r/r_{s}\right)\left(1+r/r_s\right)^2}, \label{eq:nfw}
\eeqa
where $\rho_s$ and $r_s$ are the scale density and the scale radius, respectively. These parameters 
can be condensed into one parameter, the concentration $c_{\rm vir}(M,z)$, by the use of two halo 
mass relations; namely, $M=4\pi r^3_{\rm vir} \Delta_{\rm vir}(z) \rho_{\rm crit}(z)/3$, where $r_{\rm vir}$
 is the virial radius corresponding to the overdensity criterion $\Delta_{\rm vir}(z)$ as shown, e.g., in
  Ref.~\citep{Bryan:1997dn}, and $M= \int dV \, \rho_h (\rho_s,r_s)$ with the integral performed out to 
  $r_{\rm vir}$.
In this paper, we adopt the functional form of the concentration parameter in 
Ref.~\citep{Bullock:1999he}.
The volume integral of the density squared with Eq.~(\ref{eq:nfw}) is then
\beqa
\int {\rm d}V \rho^2_{h}(r|M, z) = \frac{4\pi r^3_s \rho^2_s}{3}\left[1-\frac{1}{(1+c_{\rm vir})^3}\right]
\label {eq:rho2V}.
\eeqa
$u(k|M,z)$ and $v(k|M,z)$ represent the Fourier transform of 
density profile and density squared profile, respectively.
Both $u(k|M,z)$ and $v(k|M,z)$ are normalized so as to become unity in the limit of 
$k \rightarrow 0$. We use the Fourier transform of normalized NFW profile for $u(k|M,z)$ 
as given in Ref.~\citep{Cooray:2002dia}, and the functional form of $v(k|M,z)$ in 
Ref.~\citep{Ando:2013ff}. Finally, $b_{sh}$ is the boost factor, which is essentially equal to 
the flux multiplier $\langle \delta^2(z)\rangle$. However, in addition to the contribution 
from DM halos described in Section \ref{sec:DMann}, subhalos that reside within halos 
similarly boost the DM annihilation rate. 
We adopt the fitting formula for $b_{sh}$ provided 
by Ref.~\citep{Gao:2011rf} that includes this extra effect. Based on recent high-resolution 
dissipationless $N$-body numerical simulations, they find that
$b_{sh}=1.6\times10^{-3}\left(M/M_{\odot}\right)^{0.39}$ provides a satisfactory fit. 

The minimum halo mass $M_{\rm min}$ in Eqs.~(\ref{eq:1h}) and (\ref{eq:2h}) 
is one of the largest model uncertainties. As discussed in Section \ref{sec:DMann},
it has a large range of possibilities. For the purposes of our analysis,
we consider two cases: a conservative case with $M_{\rm min} = 10^6 M_{\odot}$
that corresponds to the typical baryonic Jeans mass \citep{Rasera:2005sa}, and an optimistic 
case with $M_{\rm min} = 10^{-6} M_{\odot}$ which is the typical free streaming 
scale for neutralino DM. In our benchmark model, the difference in $M_{\rm min}$ 
changes the amplitude of cross-correlation signal $\xi_{\delta n-\gamma_{t}}(\theta)$ 
by a factor of $\sim 10$. We regard this variation as our model uncertainty.
Namely, the 
uncertainty of our benchmark model is a factor of $\sim 10$. Note that this model
uncertainty likely dominates over
the systematic uncertainties in the Galactic diffuse template and 
those due to sample variance in our weak lensing shear measurement.

It has recently been argued that the halo profile concentration shows a 
peculiar dependence on the halo mass, and that the simple power-law extrapolation 
for concentration used in Ref.~\citep{Gao:2011rf}
results in an overestimate of the boost factor by a factor of $\sim 50$ 
depending on $M_{\rm min}$  \citep[e.g.,][]{Ng:2013xha,Sanchez-Conde:2013yxa}. 
Because most of the cross-correlation signal comes from clustering 
at large angular scales (see Figure \ref{fig:SN_xi} later in Section 
\ref{subsec:forecast}), 
our results are not strongly affected by the choice. We discuss this point 
further in detail in Appendix \ref{sec:appendixB}. 

\subsubsection{Astrophysical source contribution}

Astrophysical sources such as blazars and SFGs contribute to the EGB.
We calculate the contribution to $P_{\delta n-\kappa}(\ell)$ as 
\beqa
P_{\delta n-\kappa}(\ell)=\int \frac{{\rm d}\chi}{\chi^2} W_{g, \rm{ast}}(\chi)W_{\kappa}(\chi) P_{\delta-L}(\ell/\chi, z(\chi)),\label{eq:crosspk_astro}
\eeqa
where $W_{g, \rm{ast}}(\chi)$ represents the window function of gamma rays 
from astrophysical sources, and 
$P_{\delta-L}(k, z)$ is the three dimensional cross power spectrum of matter over density and luminosity.
The weight function $W_{g, \rm{ast}}$ is given by
\beqa
W_{g, \rm{ast}}(\chi) = \int_{E_{\rm min}}^{E_{\rm max}} 
\frac{{\rm d}E_{\gamma}}{4\pi} \ 
N_{0}(\chi)\left(\frac{E^{\prime}_{\gamma}}{E_{0}}\right)^{-\alpha}
\exp \left[ -\tau \left(E^{\prime}_\gamma, \chi \right)\right] \eta(E_\gamma),
\label{eq:Wg_ast} 
\eeqa
where 
$E_0 = 100$ MeV,
$E^{\prime}_{\gamma}=(1+z(\chi))E_{\gamma}$,
and
$N_{0}(\chi)\left(E_{\gamma}/E_{0}\right)^{-\alpha}$ 
represents the gamma-ray energy distribution of the astrophysical sources.
In modeling $P_{\delta-L}$, 
one can use a similar formalism to Eqs.~(\ref{eq:1h}) and (\ref{eq:2h}) 
but replacing the mass function $n(M,z){\rm d}M$ by the luminosity function $\Phi(L,z){\rm d}L$
\citep{Camera:2012cj}.
Assuming blazars and SFGs are well approximated as point sources, 
$P_{\delta-L}$ can be divided into two terms,
\beqa
P^{1h}_{\delta-L}(k, z)&=&\frac{1}{{\bar \rho}_{m}\langle L \rangle(\chi)}
\int_{L_{\rm min}(z)}^{L_{\rm max}(z)} {\rm d}L \ \Phi(L, z)L \ u(k|M(L),z)
\label {eq:1h_ast}\\  
P^{2h}_ {\delta-L}(k, z)&=&P^{\rm lin}(k, z) \left(\frac{1}{{\bar \rho}_{m}\langle L \rangle(\chi)}\right)
\left[\int_{M_{\rm min}} {\rm d}M \ n(M,z) b_{h}(M, z) u(k|M,z) \right] \nonumber \\
&&\times \int_{L_{\rm min}(z)}^{L_{\rm max}(z)} {\rm d}L \ \Phi(L, z)L \ b_{h}(M(L), z),
\label {eq:2h_ast}
\eeqa
where 
$\langle L \rangle (\chi)$ is the mean luminosity at $z(\chi)$
and $M(L)$ is the mass-luminosity relation of astrophysical sources.
We therefore need to set the specific functional form of $N_{0}(\chi)$, 
$\Phi(L,z)$, $M(L)$, and the power-law index of energy distribution of 
gamma-ray $\alpha$ in order to calculate
$P_{\delta n-\kappa}(\ell)$ for each astrophysical source.

For the gamma-ray luminosity function of blazars, 
we adopt the luminosity-dependent density evolution model 
\citep{Narumoto:2006qg, Ando:2006mt} 
with parameters in Ref.~\citep{Ando:2013ff}.
We set the power law index $\alpha$ for blazars to be 2.4, 
which is consistent with the spectra of resolved blazars.
The gamma-ray luminosity of blazars is evaluated as $\nu L_{\nu}$ at 100 MeV.
In this case, $N_{0}$ is given by $\langle L \rangle/E_{0}^2$.
We adopt the mass-luminosity relation 
$
M(L) = 10^{11.3} M_{\odot} \left(L/10^{44.7} \ {\rm erg}\ {\rm s}^{-1}\right)^{1.7}
$ 
that yields the desired bias of blazer host halos \citep{Ando:2006mt}.
We assume that there are no blazars fainter 
than the luminosity $L_{\rm min}=10^{42}$ erg ${\rm s}^{-1}$ at any redshift.
In estimating $L_{\rm max}(z)$, we assume 
a blazar can be resolved if the gamma-ray flux $F$ at $E > 100$ MeV 
is larger than $2\times10^{-9} \ {\rm cm}^{-2} \ {\rm s}^{-1}$.

For the gamma-ray luminosity function of SFGs, 
we use the tight correlation between the infrared (IR) luminosity and
the gamma-ray luminosity \citep{Ackermann:2012vca}, and
use the observed IR luminosity function \citep{Rodighiero:2009up}.
We define gamma-ray luminosity in the energy range between 0.1 GeV and 100 GeV,
and we assume a power-law spectrum with index $\alpha = 2.7$ for SFGs. 
This leads to 
$N_{0}(\chi) = (\langle L \rangle/E_{0}^2)(\alpha-2)/(1+z(\chi))^{2-\alpha}$
so that the mean luminosity is obtained as
$\langle L \rangle=\int{\rm d}E_{\gamma} \ E_{\gamma}N_{0}(\chi)\left(E_{\gamma}/E_{0}\right)^{-\alpha}$
with the integral performed from $(1+z)E_{0}$ to $(1+z)E_{1}$,
where $E_{0}=100$ MeV and $E_{1} = 100$ GeV.
We use the mass-luminosity relation for 
SFGs,
$
M(L) = 10^{12} M_{\odot}\left(L/10^{39} \ {\rm erg}\ {\rm s}^{-1}\right)^{0.5}
$
that is calibrated by the Milky Way properties
\citep{Camera:2012cj}.
The minimum luminosity is set to $10^{30}$ ${\rm erg}\ {\rm s}^{-1}$ at any redshift, 
while the maximum luminosity is estimated in the same way as in the case of blazars.

\begin{figure}[!t]
\begin{center}
       \includegraphics[clip, width=0.5\columnwidth]{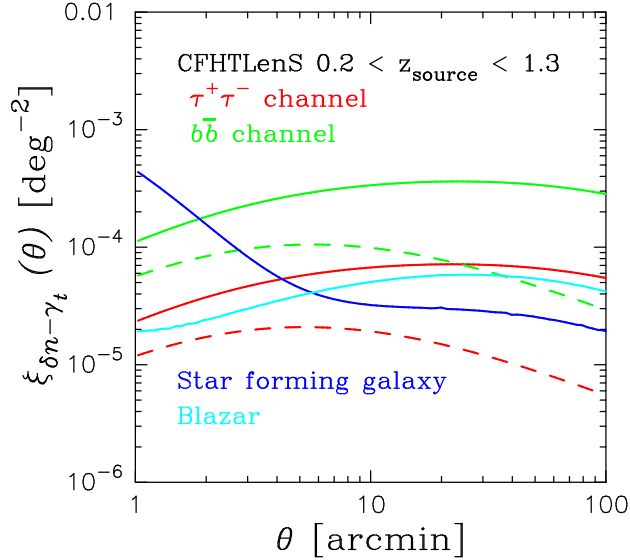}
       
    \caption{
    \label{fig:xi_dm}
   The expected cross-correlation signals of cosmic shear and important components of the EGB:  
   from SFG (blue), blazers (cyan), and DM annihilation. For the latter, we show the signal from 
   a 100 GeV DM particle with annihilation cross section 
   $\langle \sigma v \rangle=3 \times 10^{-26} \ {\rm cm}^3 \ {\rm s}^{-1}$
   and annihilation channels $\tau^{+}\tau^{-}$ (red) and $b {\bar b}$ (green).
   Furthermore we consider two values for the minimum halo mass; 
   $M_{\rm min} = 10^{-6} M_\odot$ (solid)
   and $M_{\rm min} = 10^{6}  M_\odot$ (dashed). 
  } 
    \end{center}
\end{figure}

Figure \ref{fig:xi_dm} shows our benchmark model of cross-correlation signals 
in the case of DM annihilation 
with $m_{dm}=100$ GeV and $\langle \sigma v \rangle=3 \times 10^{-26} \ {\rm cm}^3 \ {\rm s}^{-1}$.
In this figure, the results for two annihilation channels are shown, 
the $\tau^{+}\tau^{-}$ channel (red lines) 
and the $b {\bar b}$ channel (green lines). 
We show the level of model uncertainty due to the minimum halo mass $M_{\rm min}$
by plotting both the optimistic case with $M_{\rm min} = 10^{-6} M_{\odot}$ (solid lines) 
and the conservative case 
with $M_{\rm min} = 10^{6} M_{\odot}$ (dashed lines).
The figure clearly shows the sensitivity of the results on $M_{\rm min}$ and 
the different annihilation channels. The blue and cyan line in 
figure \ref{fig:xi_dm} 
show the cross-correlation signals of cosmic shear 
and EGB contributed by unresolved SFGs and blazars, respectively.
Clearly, the contribution from astrophysical sources 
can be significant at all angular scales.
We note that our adopted model of blazars is different from the one in the 
previous work of Ref.~\citep{Camera:2012cj}.
Our model reproduces the observed flux counts of resolved blazars, 
whereas the model in Ref.~\citep{Camera:2012cj} is aimed at reproducing the flux counts 
as well as the anisotropy of the EGB \citep{Harding:2012gk}.
The main difference lies in the faint slope of the gamma-ray luminosity function.
Overall, our model predicts a larger contribution from blazers to the EGB intensity 
than the model of Ref.~\citep{Camera:2012cj} by a factor of $\sim 10$.
The large model-difference unfortunately limits the extent to which we can
subtract astrophysical contributions. 
In this paper, we first examine the case where DM annihilation is the sole contributor 
to the cross-correlation signal. Our analysis under this assumption should 
provide a conservative constraint on DM annihilation, 
because the astrophysical sources are expected to 
yield positive cross-correlation signals unless
they are distributed in an anti-correlated manner
with respect the underlying DM density field.
Furthermore, we find that the 
statistical error in the current dataset is larger than the expected cross-
correlation signals due to astrophysical sources. Therefore, the final result is 
not strongly dependent on the details of the models for the astrophysical sources.

\subsubsection{Point spread function}

The observed number of EGB photons along a line of sight ${\bd \theta}$
is expressed by the convolution of the underlying number of EGB photons 
with the PSF of the detector,
\beqa
\delta n^{\rm obs}({\bd \theta}) 
=\int {\rm d}^2\theta^{\prime}\, 
W_{\rm PSF}({\bd \theta}-{\bd \theta}^{\prime})
\delta n({\bd \theta}^{\prime}),
\eeqa
where $\delta n^{\rm obs}$ is the observed number of EGB photons and 
$W_{\rm PSF}$ is the PSF. This causes 
an additional scale dependence of the weight function of EGB counts 
in Eqs.~(\ref{eq:Wg_dm}) and (\ref{eq:Wg_ast}).
Considering the energy dependence of the PSF,
the scale-dependent weight function is given by
\beqa
W_{g}(\chi) \rightarrow W_{g}(\chi, \ell) 
&=& \int_{E_{\gamma, \rm min}}^{E_{\gamma, \rm max}} {\rm d}E_{\gamma}
 \ \frac{\langle \sigma v \rangle}{8\pi}
\left(\frac{{\bar \rho}_{\rm dm, 0}}{m_{\rm dm}}\right)^2
\left[1+z(\chi)\right]^3 \frac{{\rm d}N_{\gamma}}{{\rm d}E_{\gamma}}
\Bigg|_{E^{\prime}_{\gamma}} \nonumber \\
&&
\, \, \, \, \, \, \, \, \, \, \, \, 
\, \, \, \, \, \, \, \, \, \, \, \, 
\, \, \, \, \, \, \, \, \, \, \, \, 
\times \exp \left[ -\tau \left(E^{\prime}_{\gamma}, \chi \right)\right] 
\eta(E_{\gamma}) \tilde{W}_{\rm PSF}(\ell, E_{\gamma}), 
\label{eq:Wg_dm_psf} \\
W_{g, \rm{ast}}(\chi) \rightarrow W_{g, \rm{ast}}(\chi, \ell)
&=& \int_{E_{\rm min}}^{E_{\rm max}} 
\frac{{\rm d}E_{\gamma}}{4\pi} \ 
N_{0}(\chi)\left(\frac{E^{\prime}_{\gamma}}{E_{0}}\right)^{-\alpha} \nonumber \\
&&
\, \, \, \, \, \, \, \, \, \, \, \, 
\, \, \, \, \, \, \, \, \, \, \, \, 
\, \, \, \, \, \, \, \, \, \, \, \, 
\times \exp \left[ -\tau \left(E^{\prime}_\gamma, \chi \right)\right] 
\eta(E_\gamma)\tilde{W}_{\rm PSF}(\ell, E_{\gamma}),
\label{eq:Wg_ast_psf} 
\eeqa
where $\tilde{W}_{\rm PSF}(\ell, E_{\gamma})$ is the fourier transform of the PSF.

\begin{figure}[!t]
\begin{center}
       \includegraphics[clip, width=0.5\columnwidth]
       {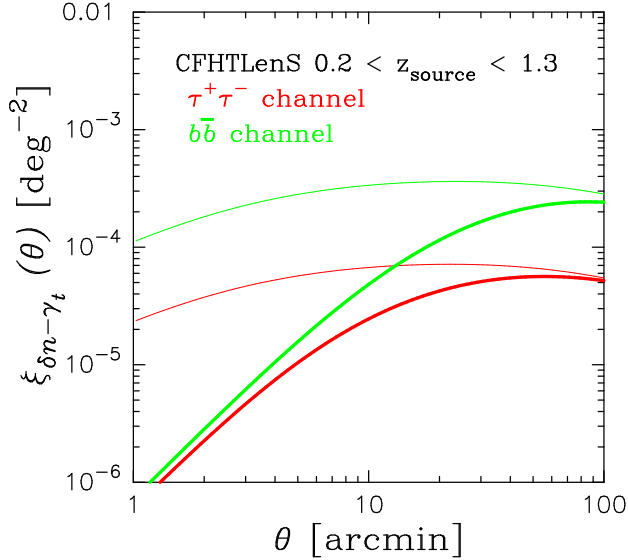}
       
    \caption{
    \label{fig:xi_dm_psf}
   The smoothing effect due to the PSF on the
   cross-correlation signals of cosmic shear and EGB. 
   The thin lines represent the original expected signal as in Figure \ref{fig:xi_dm}:
   annihilation of a 100 GeV mass DM with annihilation cross section 
   $\langle \sigma v \rangle=3 \times 10^{-26} \ {\rm cm}^3 \ {\rm s}^{-1}$
   and minimum halo mass $M_{\rm min} = 10^{-6} M_\odot$; red and green
   lines are for the $\tau^{+}\tau^{-}$ and $b {\bar b}$ channel, respectively, 
   while the thick lines show the signal with smoothing due to the PSF.   
   } 
    \end{center}
\end{figure}

In the case of {\it Fermi}-LAT, the PSF is modeled using the following functional form \citep{Ackermann:2012kna}:
\beqa
W_{\rm PSF}(\theta, E_{\gamma})
&=&
A(E_{\gamma}) \left[f_{\rm core}K(x, \sigma_{\rm core}, \gamma_{\rm core})
+(1-f_{\rm core})K(x, \sigma_{\rm tail}, \gamma_{\rm tail})\right], 
\label{eq:fermi_psf} \\
f_{\rm core}&=&\frac{1}{1+N_{\rm tail}\sigma_{\rm tail}^2/\sigma_{\rm core}^2}, \\
K(x, \sigma, \gamma) 
&=&
\frac{1}{2\pi\sigma^2}\left(1-\frac{1}{\gamma}\right)
\left[1+\frac{1}{2\gamma}\frac{x^2}{\sigma^2}\right]^{-\gamma},
\eeqa
where $x$ is a scaled-angular deviation defined by 
$x=\theta/S_{\rm P}(E_{\gamma})$ 
and 
$A(E_{\gamma})$ is the normalization factor 
such that $\int {\rm d}^{2}\theta \, W_{\rm PSF}(\theta, E_{\gamma})=1$.
The scale factor $S_{\rm P}(E_{\gamma})$ is \citep{Ackermann:2012kna},
\beqa
S_{\rm P}(E_{\gamma})=\sqrt{\left[c_{0}\left(\frac{E_{\gamma}}{100\, {\rm MeV}}\right)^{-\beta}\right]^2+c_{1}^2},
\eeqa
and the normalization is given by $A(E_{\gamma}) = [S_{\rm P}(E_{\gamma})]^2$.
In the present paper, we adopt the parameters estimated in the latest in-flight PSF
for {\tt ULTRACLEAN} photons \footnote{made publicly available at \\ http://fermi.gsfc.nasa.gov/ssc/data/analysis/documentation/Cicerone/Cicerone\_LAT\_IRFs/IRF\_PSF.html},
i.e., $c_{0}=3.16$ deg and $c_{1}=0.034$ deg for front-converting events,
and 
$c_{0}=5.32$ deg and $c_{1}=0.096$ deg for back-converting events,
along with 
$\beta = 0.8$, $N_{\rm tail} =0.08639$, $\sigma_{\rm core}=0.5399$, $\sigma_{\rm tail}=1.063$,
$\gamma_{\rm core}=2.631$, and $\gamma_{\rm tail}=2.932$ for both events \citep{Ackermann:2012kna}.

Using the specific functional form shown in Eq.~(\ref{eq:fermi_psf}),
we estimate the effect of the PSF on the cross-correlation analysis.
In Figure \ref{fig:xi_dm_psf}, we consider the cross-correlation signal due to 
the annihilation of DM with $m_{\rm dm}= 100$ GeV and 
$\langle \sigma v \rangle=3 \times 10^{-26} \ {\rm cm}^3 \ {\rm s}^{-1}$.
To account for the PSF, 
we first calculate the cross-correlation signals with the scale-dependent weight function
in Eqs.~(\ref{eq:Wg_dm_psf}) and (\ref{eq:Wg_ast_psf}) 
for front- and back-converting events, respectively.
We then average these two signals at a given angular separation
assuming the number of front-converting events is equal to that of back-converting
events.
Clearly, the smoothing effect significantly affects the cross-correlation signal especially at 
smaller angular scales than the typical size of the PSF, i.e. $\sim 50$ arcmin.
We also expect that the pixelization effect would be unimportant in our analysis,
because the pixel size is smaller than the size of the PSF (12 arcmin).

\section{\label{sec:res}RESULT}

\begin{figure}[!t]
\begin{center}
       \includegraphics[clip, width=0.45\columnwidth]{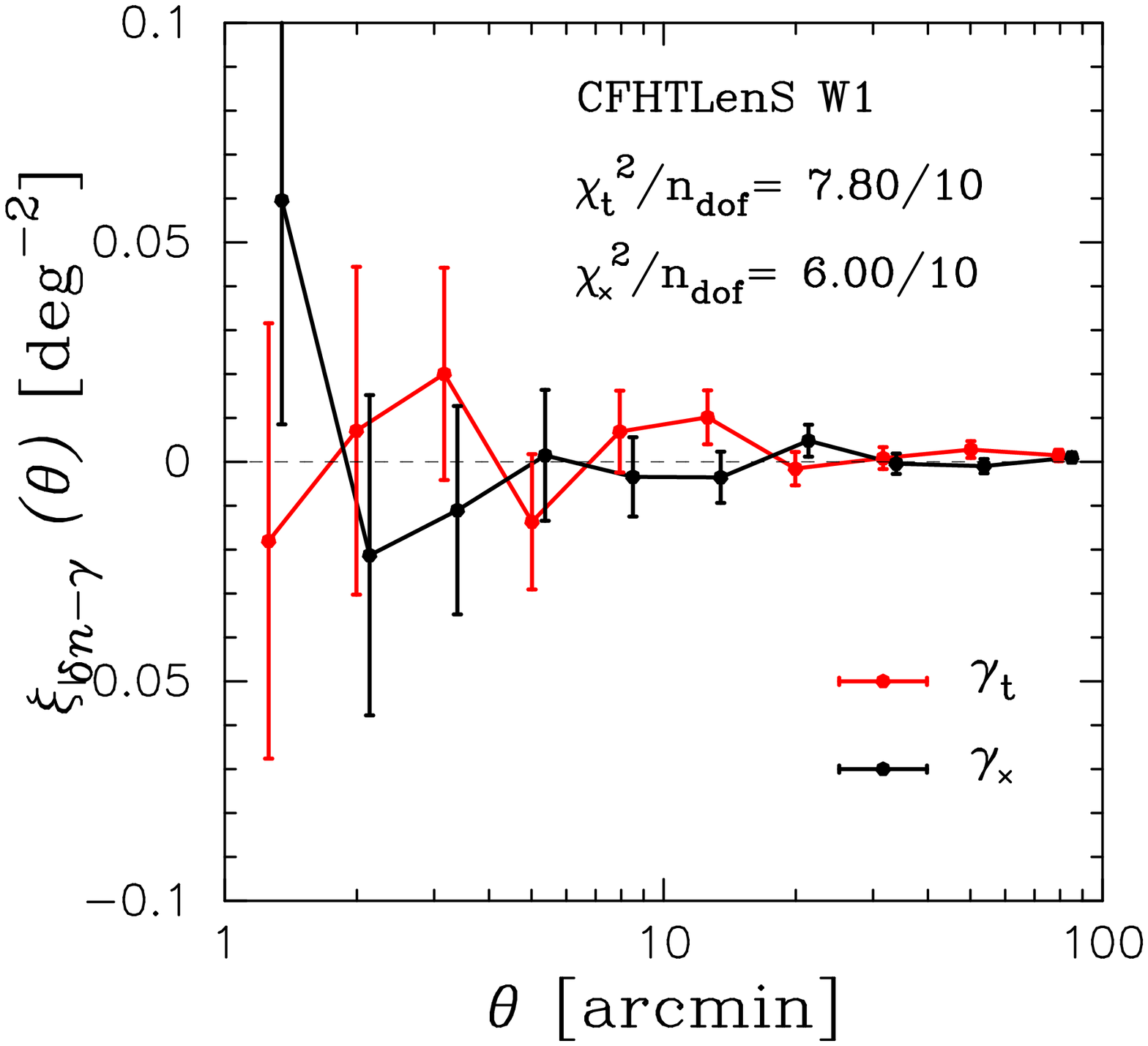}
       \includegraphics[clip, width=0.45\columnwidth]{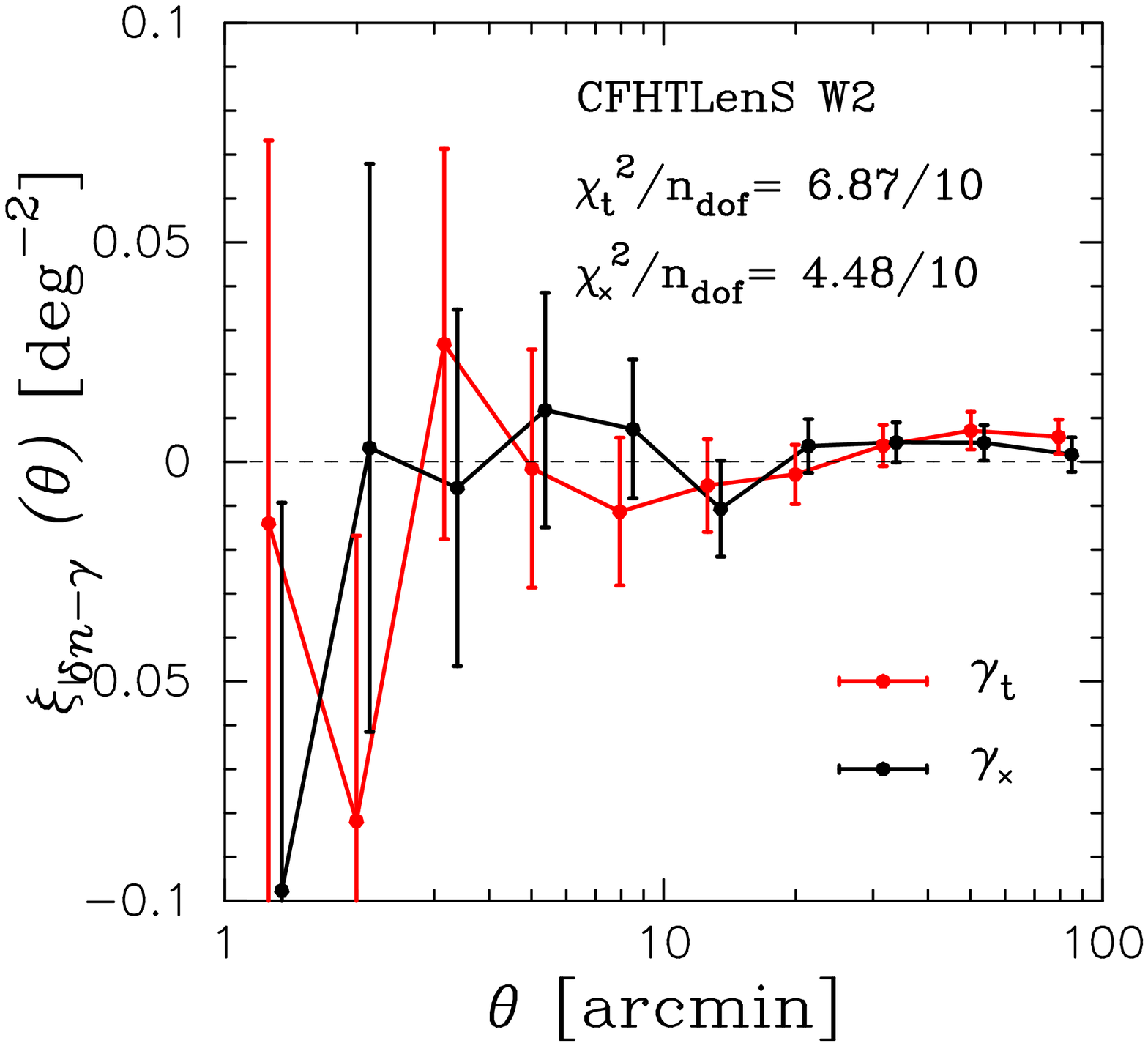}
       \includegraphics[clip, width=0.45\columnwidth]{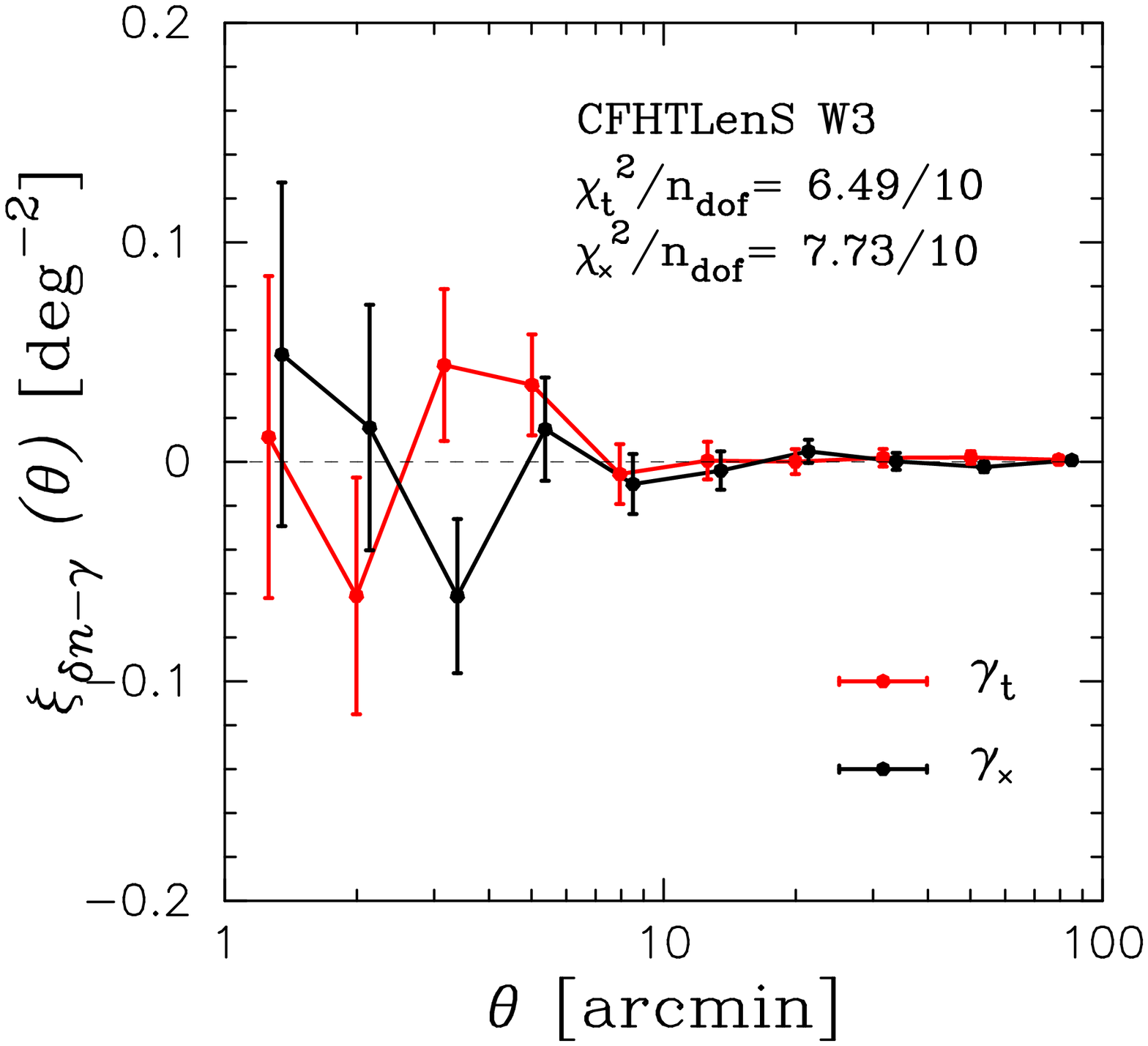}
       \includegraphics[clip, width=0.45\columnwidth]{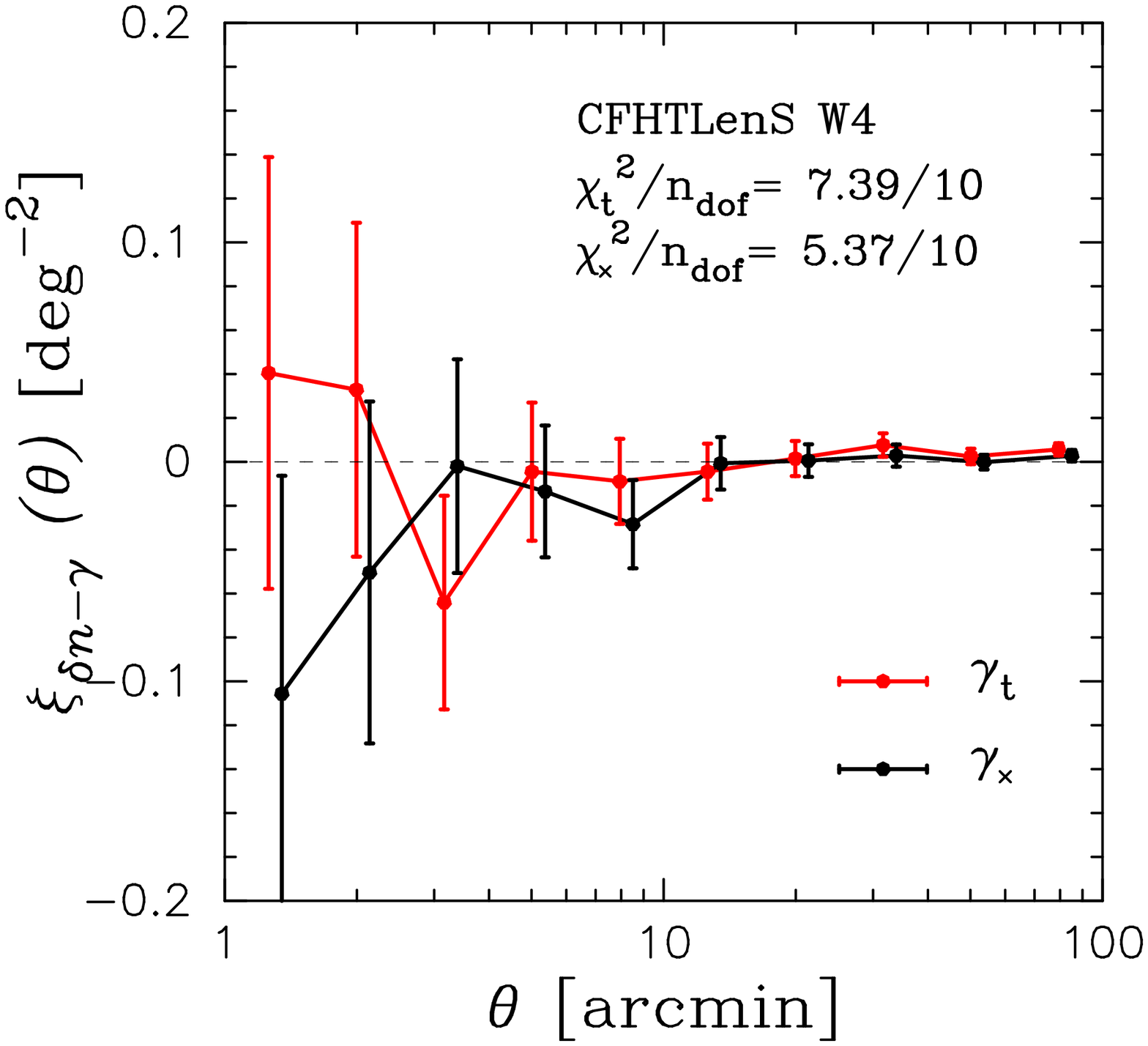}
    \caption{
    \label{fig:xi_1G500G}
    The cross-correlation signal of cosmic shear and the EGB. Each panel corresponds to 
    each of the CFHTLenS patches W1-W4. 
    The red points show the result using tangential shear $\gamma_{+}$, while the black points are
    for $\gamma_{\times}$.
    The error bars indicate the standard deviation estimated from our 500 randomized shear catalogues
    and 500 randomized photon count maps.
	     } 
    \end{center}
\end{figure}

We present the measurement of the cross-correlation signals 
of the cosmic shear and the EGB and discuss the implications.
Figure \ref{fig:xi_1G500G} shows the cross-correlation signals obtained
for each CFHTLenS patch.
In each panel of Figure \ref{fig:xi_1G500G},
we also show the cross-correlation using another component of weak lensing 
shear that is rotated $45^\circ$ from the tangential shear component.
We refer to this component as $\gamma_{\times}$.
In practice, $\gamma_{\times}$ is often used as an indicator of systematics 
in the shape measurement.
In the case of $\it perfect$ shape measurement and no intrinsic alignment, the
correlation signal with $\gamma_{\times}$ should vanish statistically.
In order to quantify the significance of the measured 
cross-correlation signals with respect to the statistical error, 
we use the $\chi^2$ statistics defined by
\beqa
\chi^2 
= \sum_{i,j}\xi_{\delta n-\gamma_{t}}(\theta_{i})C^{-1}_{ij}\xi_{\delta n-\gamma_{t}}(\theta_{j}),
\eeqa
where ${\bd C}^{-1}$ denotes the inverse covariance matrix 
estimated from the randomized realization shown in Section~\ref{subset:analysis}.
In our analysis, the number of deg of freedom is 10.
The resulting values of $\chi^2 / n_{\rm dof}$ for $\gamma_{t}$
and for $\gamma_{\times}$ are shown in each panel. 
The result is consistent with null detection in each 
CFHTLenS patch. 
We confirm that the combined 
four field together is also 
consistent with null detection
($\chi^2/n_{\rm dof}=[7.80+6.87+6.49+7.39]/40=28.55/40$ in total).

\begin{figure}[!t]
\begin{center}
       \includegraphics[clip, width=0.45\columnwidth]
       {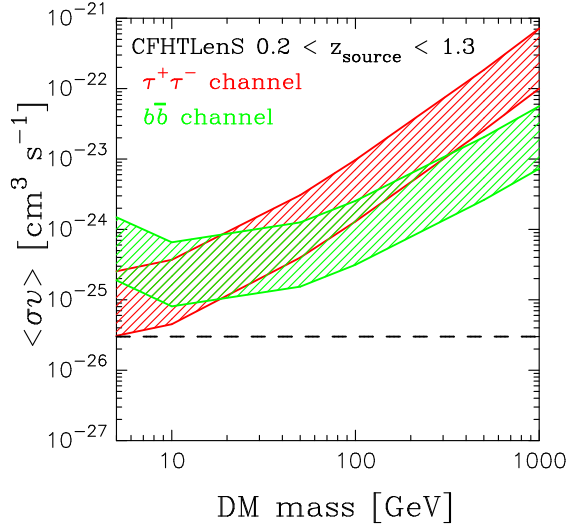}
    \caption{
     \label{fig:sigv_m_CFHT}
   The 68 \% confidence level upper limits on $\langle \sigma v \rangle$ as a function of 
   DM mass. The red shaded region shows the upper bound for the $\tau^{+}\tau^{-}$ 
   channel and the green region is for the $b {\bar b}$ channel.
   Note that the widths of the shaded regions indicate the model uncertainty:
   for each shaded region, the upper curve is derived by our benchmark model with 
   $M_{\rm min}
   = 10^6 M_{\odot}$ and the lower curve is obtained
   from the model with $M_{\rm min} = 10^{-6} M_{\odot}$.
	     } 
    \end{center}
\end{figure}

We are now able to use the null detection of the cross-correlation to place 
constraints on the DM annihilation cross-section.
For this purpose, we use the maximum Likelihood analysis.
We assume that the data vector ${\bd D}$ is well approximated 
by the multivariate Gaussian distribution with covariance ${\bd C}$.
In this case, $\chi^2$ statistics (log-likelihood) is given by
\beqa
\chi^2({\bd p}) = \sum_{i,j}(D_{i}-\mu_{i}({\bd p}))C^{-1}_{ij}(D_{j}-\mu_{j}({\bd p})), \label{eq:logL}
\eeqa
where ${\bd \mu}({\bd p})$ is the theoretical prediction 
as a function of parameters of interest.
In this paper, we use the halo model approach shown
in Section \ref{subsec:halomodel} to calculate the theoretical prediction.
For parameters of interest ${\bd p}$,  
we simply consider the DM particle mass and the annihilation 
cross-section, $m_{\rm dm}$ and $\langle \sigma v \rangle$\footnote{
Strictly speaking, we need to consider other parameters associated with the model 
of substructure within DM haloes.
These are, for example, the concentration parameter $c_{\rm vir}$ of host halo, 
subhalo density profile and subhalo mass function.
Although we do not include these parameters explicitly in our analysis, 
we explore the overall effect by considering 
two cases with the different minimum halo mass 
$M_{\rm min}$ as the most important $\it effective$ uncertainty of our benchmark model.
}
The data vector ${\bd D}$ consists of the measured cross-correlation 
signals with the range of $\theta=[1, 100]$ arcmin
as 
\beqa
D_{i} = \{ \xi_{\delta n-\gamma_{t}}(\theta_{1}), \xi_{\delta n-\gamma_{t}}(\theta_{2}),..., \xi_{\delta n-\gamma_{t}}(\theta_{10}) \},
\eeqa
where $\theta_{i}$ is the $i$-th bin of angular separation.
The inverse covariance matrix ${\bd C}^{-1}$ includes 
the statistical error of the shape measurement 
and the photon Poisson error.
In our likelihood analysis, we assume that 
the four CFHTLenS patches are independent of each other.
With this assumption, the total log-likelihood is given by 
the summation of Eq.~(\ref{eq:logL}) in each CFHTLenS patch.
In order to constrain $m_{\rm dm}$ and $\langle \sigma v \rangle$,
we consider the 68 \% confidence level of posterior distribution function of parameters.
This is given by the contour line in the two dimensional space 
($m_{\rm dm}$ and $\langle \sigma v \rangle$), which is defined as 
\beqa
\Delta \chi^2({\bd p}) = \chi^2({\bd p})-\chi^2({\bd \mu}=0)=2.30.
\eeqa

As discussed in Section \ref{subsec:halomodel}, the choice of the minimum halo mass
affects the theoretical predictions by a factor of about ten.
We therefore derive constraints based on the optimistic case with
$M_{\rm min }=10^{-6} M_{\odot}$ and on the conservative case with
$M_{\rm min }=10^{6} M_{\odot}$.

Figure \ref{fig:sigv_m_CFHT} shows the result of our likelihood analysis on the
DM parameter space $m_{\rm dm}$ and $\langle \sigma v \rangle$.
We plot the constraints for two representative particle physics 
model, the $\tau^{+}\tau^{-}$ channel and  the $b \bar{b}$ 
channel. We also show the results for the two choices of $M_{\rm min}$. 
The constraint for the small $M_{\rm min}$ is
significantly stronger, as expected. At low DM mass,
the annihilation cross-section is more severely constrained 
for the $\tau^+ \tau^-$ channel,
because of its harder gamma-ray spectra that contribute photons at sensitive 
energies than for the $b \bar{b}$ channel of the same DM mass. 
For reference, the horizontal
dashed line indicates the canonical cross section of
$\langle \sigma v \rangle = 3 \times 10^{-26}\, {\rm cm}^{3}\, {\rm s}^{-1}$
for a thermally produced DM. 

\subsection{\label{subsec:forecast}Future forecast}

Future weak lensing surveys are aimed at measuring cosmic shear
over a wide area of more than a thousand square degs. 
Such observational programs include the
Subaru Hyper Suprime-Cam (HSC) \footnotemark[1],  
the Dark Energy Survey (DES) \footnotemark[2], 
and the Large Synoptic Survey Telescope (LSST) \footnotemark[3].
\footnotetext[1]{\rm{http://www.naoj.org/Projects/HSC/j\_index.html}} 
\footnotetext[2]{\rm{http://www.darkenergysurvey.org/}}
\footnotetext[3]{\rm{http://www.lsst.org/lsst/}}
It is interesting to explore the discovery potential of the
upcoming cosmology surveys in terms of the DM particle properties. 
In this section, we consider two of these wide surveys with an area coverage 
of 1400 ${\rm deg}^2$ (HSC)
and 20000 ${\rm deg}^2$ (LSST),
by simply scaling the covariance matrix by a factor 
of $154/1400$ or $154/20000$, respectively. Assuming the same number density 
and redshift distribution of source galaxies as in the CFHTLenS,
the expected constraints can be scaled by the effective survey area.
The result suggests that the upper limit will be improved by a factor of $\sqrt{1400/154}\sim3$ 
for HSC and by a factor of $\sqrt{20000/154}\sim11$ for LSST.
In particular, for a 100 GeV DM, the upper limit of $\langle \sigma v \rangle$ with 
68 \% confidence level could reach
$2.7-22.2\times10^{-26} \ {\rm cm}^{3} \ {\rm s}^{-1}$ for $b \bar{b}$ channel
and
$1.1-8.51\times10^{-25} \ {\rm cm}^{3} \ {\rm s}^{-1}$ for $\tau^{+}\tau^{-}$ channel
in the case of the LSST-like survey.
It will be important to include the uncertainty in the model template of 
galactic emission and also the sampling variance that is neglected in this paper.
Then we will be able to derive 
robust and complementary probes of DM annihilation from
the cross-correlation signal of cosmic shear and EGB.

\begin{figure}[!t]
\begin{center}
       \includegraphics[clip, width=0.45\columnwidth]
       {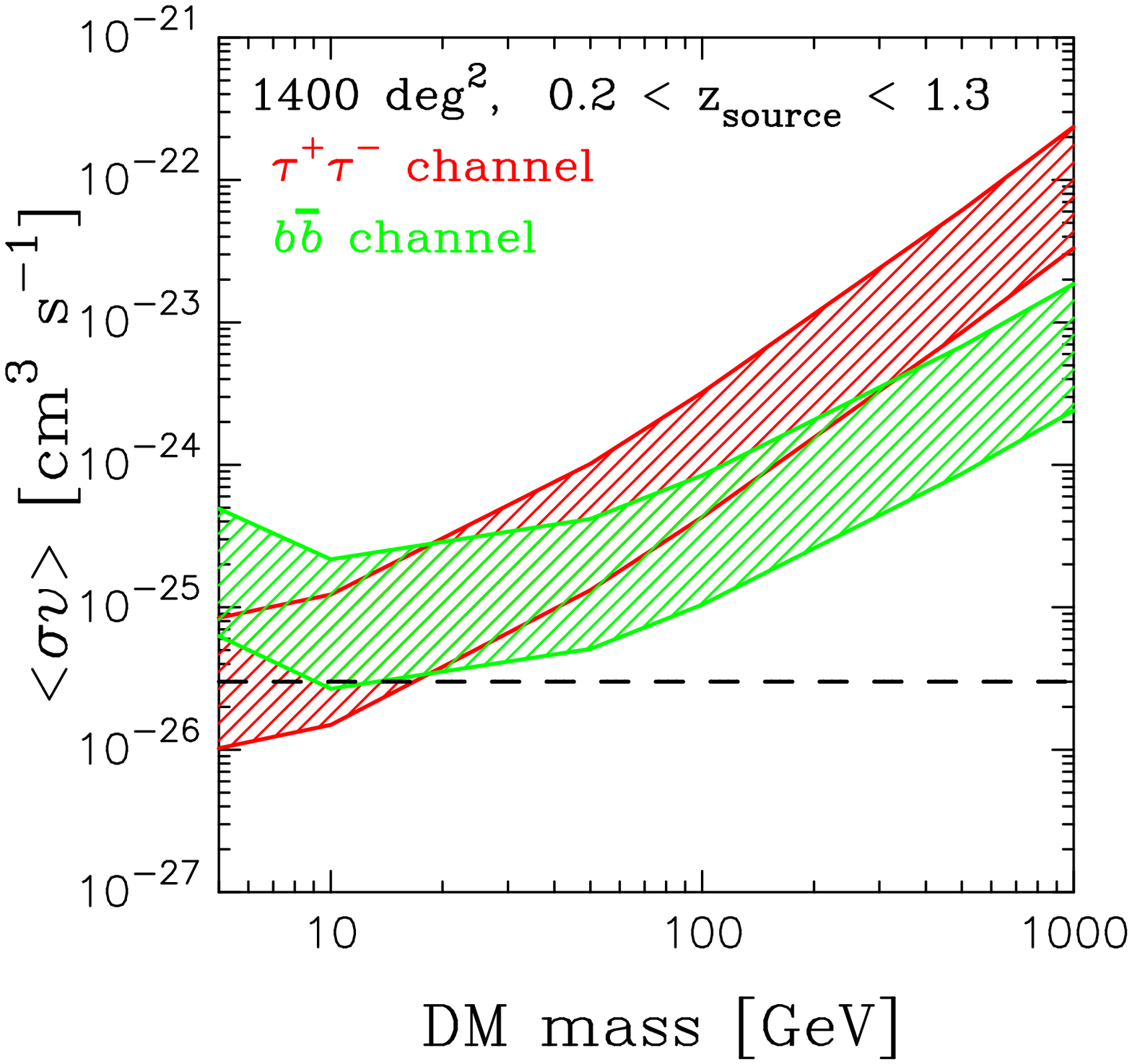}
       \includegraphics[clip, width=0.45\columnwidth]
       {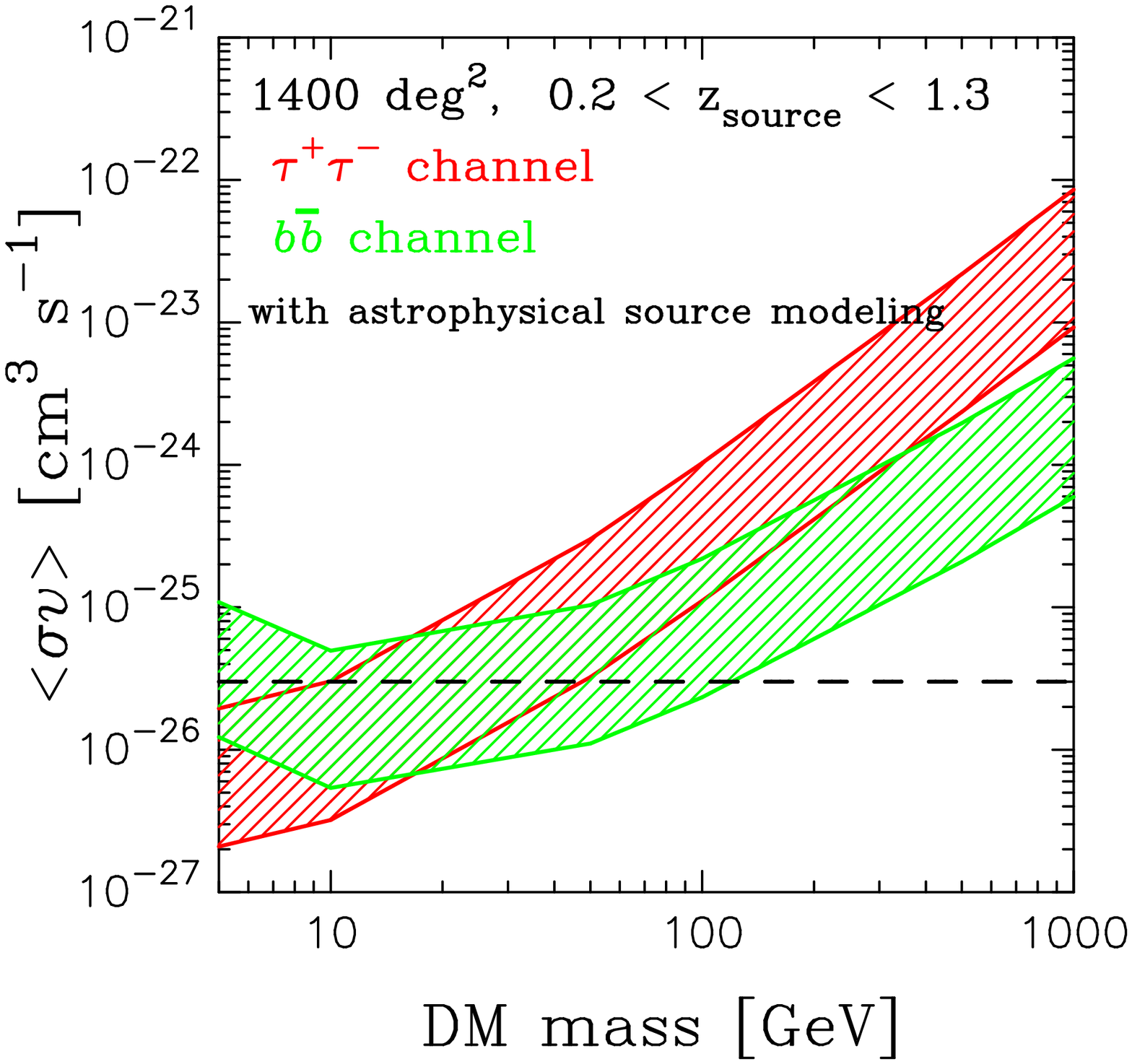}
    \caption{
     \label{fig:sigv_m_HSC}
   We plot the expected 68 \% confidence level upper limit on $\langle \sigma v \rangle$ 
   as a function of the DM mass for upcoming surveys. We show the case with a sky 
   coverage of survey area 1400 ${\rm deg}^2$.
   The red shaded region shows the expected upper limit for the $\tau^{+}\tau^{-}$ channel
   and the green one for the $b {\bar b}$ channel.
   The left panel shows that the conservative case assuming the DM annihilation 
   contribution only, while the right panel shows the optimistic case taking into account 
   astrophysical sources.
  } 
    \end{center}
\end{figure}

As shown in Figure \ref{fig:xi_dm}, 
the expected cross-correlation of astrophysical sources are 
comparable to the DM annihilation signal with
$m_{\rm dm}=100$ GeV and $\langle \sigma v \rangle=3 \times 10^{-26} \ {\rm cm}^3 \ {\rm s}^{-1}$.
Thus it will be even more important to accurately take into account of 
the contribution of astrophysical sources such as blazars and SFG for future surveys.
We thus include the contribution from the astrophysical sources on the assumption
that the contribution of blazars and SFGs can be estimated 
as in our benchmark model described in \ref{subsec:halomodel}.
The sum of the three contributions is given by
\beqa
\xi_{\delta n-\gamma_{t}}(\theta) = 
\xi_{\delta n-\gamma_{t}}^{\rm dm}(\theta|m_{\rm dm}, \langle \sigma v \rangle)
+\xi_{\delta n-\gamma_{t}}^{\rm blazer}(\theta)
+\xi_{\delta n-\gamma_{t}}^{\rm SFG}(\theta).
\eeqa
Using this as a theoretical model template, 
we perform the likelihood analysis 
to make forecast for DM constraints.
For simplicity, we assume that
the observed cross-correlation is identical to the one of the CFHTLenS W1 patch
but that the covariance matrix can be scaled by the survey area.
The expected constraint from the HSC-like survey is shown in Figure \ref{fig:sigv_m_HSC}.
The left panel shows the conservative case with no contribution from the
astrophysical sources whereas
the right panel shows the case with including the astrophysical sources.
With the astrophysical sources in the model prediction, 
we can place tighter upper bound by $\sim 40-70 \%$ 
for the sky coverage of 1400 ${\rm deg}^2$.
It is clearly important to treat the contribution from the astrophysical sources 
carefully
for future wide-field surveys.

\begin{figure}[!t]
\begin{center}
       \includegraphics[clip, width=0.45\columnwidth]{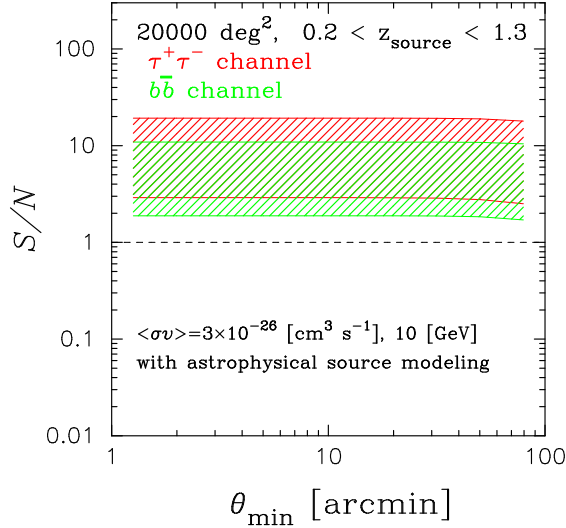}
    \caption{
     \label{fig:SN_xi}
   The cumulative signal-to-noise ratio for the cross-correlation of cosmic shear and 
   the EGB. We show the case with a sky coverage of survey area 20000 ${\rm deg}^2$,
   i.e., a LSST like survey.
   The red shaded region shows the signal-to-noise ratio for the $\tau^{+}\tau^{-}$ channel
   and the green one for the $b {\bar b}$ channel.
   We consider the sum of the DM annihilation contribution 
   of a 10 GeV mass DM
   and the astrophysical sources for these plots. 
     } 
    \end{center}
\end{figure}

We further study information content in the cross-correlation signal of cosmic shear and EGB.
An important quantity is the cumulative signal-to-noise ratio $S/N$, which is defined by
\beqa
\left(S/N\right)^2 = \sum_{i,j}\mu_{i}({\bd p})C^{-1}_{ij}\mu_{j}({\bd p}).
\eeqa
In order to calculate $S/N$, we consider DM models with
$\langle \sigma v \rangle = 3 \times 10^{-26} \ {\rm cm}^3 \ {\rm s}^{-1}$ 
for a 10 GeV and 100 GeV dark matter
and use the covariance matrix estimated by the randomized method shown in Section~\ref{subset:analysis}.

Figure \ref{fig:SN_xi} shows the $S/N$ 
as a function of the minimum angular scale included in the cross-correlation analysis.
In this figure, we consider the annihilation signal of a 10 GeV DM particle and
we set the maximum angular scale to 100 arcmin.
Large-scale cross-correlations determine the information 
content, and including data at small angular scales does not
improve the significance.
The same can be said of a 100 GeV DM particle.
This is simply because we can not extract information from cross-correlations
on scales smaller than the size of the gamma-ray PSF.
At large angular scales, $\theta \sim 100 \ {\rm arcmin}$, 
the signals are mainly contributed by the DM annihilation.
We expect that the cross-correlation analysis with
upcoming survey with a large sky converge of $\sim 1000 \ {\rm deg}^2$ 
will be a powerful probe of dark matter annihilation.
We also discuss the detectability of the cross-correlation signal with upcoming lensing surveys.
In our benchmark model, 
the $S/N$ is almost proportional to $\langle \sigma v \rangle$
because the DM contribution dominates over astrophysical contributions.
We can thus ${\it detect}$ at a 3-$\sigma$ confidence level the DM signature with
$\langle \sigma v \rangle \simeq 3 \times 10^{-26} \ {\rm cm}^3 \ {\rm s}^{-1}$ 
for a 10 GeV dark matter
and 
$\langle \sigma v \rangle \simeq 1 \times 10^{-25} \ {\rm cm}^3 \ {\rm s}^{-1}$ 
for a 100 GeV dark matter
in a LSST-like survey.
It is important to note that 
$S/N$ will likely increase significantly if cross-correlations 
at very large angular scales ($\simgt 100$ arcmin) are included.
In the present paper,
the statistical error estimated from the real dataset
is limited to the range of 1--100 arcmin.
However, for upcoming wide-field surveys, we can measure the cross-correlation 
signal to much larger angular scales
where the smoothing effect due to PSF is unimportant.
To estimate the expected value of $S/N$ in upcoming surveys, 
one would need 
mock weak lensing catalogues and gamma-ray photon maps 
with a sky coverage of $\simgt 1000$ squared degs.
This is along the line of our ongoing study 
using a large set of cosmological simulations in combination 
with actual Fermi all-sky observations.
It is important to note that our method shown in the present paper 
probes the DM signature at cosmological scales,
and thus is complementary to DM searches in local galaxies.

\section{\label{sec:con}CONCLUSION AND DISCUSSION}

We have performed, for the first time, cross-correlation analysis of cosmic shear and the EGB 
using observational data from the CFHTLenS and the $\it Fermi$ satellite. 
For the 154 square-degs sky coverage, the measured cross-correlation signal is 
consistent with null detection. Using theoretical models based on large-scale DM 
structure formation, 
we have estimated the statistical error from real data together with a large set of mock
observations, 
and have placed 
constraints on the DM annihilation cross section. We have considered different DM
annihilation channels and varied the minimum mass of DM halos. 
The derived constraint is  
$\langle \sigma v \rangle < 10^{-25}-10^{-24} \ {\rm cm}^{3} \ {\rm s}^{-1}$ for a $100$ 
GeV DM, depending on the assumed parameters and annihilation channel. The 
constraint improves for smaller DM mass. 

Recent analyses of the Fermi observations of dwarf 
galaxies \citep{GeringerSameth:2011iw,Ackermann:2011wa,Ackermann:2013yva} 
provide stronger constraints for DM annihilation. 
However, our constraints are derived using a completely different 
statistical method,
based on the cross-correlation of the EGB and cosmic shear.
The EGB \emph{intensity} has been used to constrain the DM contribution,
most recently by modeling and removing the astrophysical sources to obtain strong
limits \citep{Cholis:2013ena}. 
Our limits compete favorably 
with the constraints of Ref.~\citep{Ando:2012vu} that use galaxy clusters
and those of Ref.~\citep{Ando:2013ff} that use anisotropies of the EGB.
Given the range of potential DM signals in the literature and a broad range of potential 
particle candidates, complementary probes are critical to cast a wide net for DM 
signals and constraints. For example, recently a $\sim $GeV excess has been 
claimed towards the Galactic center whose spectral shape, normalization, and 
spatial morphology can all be explained by the annihilation of $10$ GeV ($40$ GeV)
mass DM to $\tau^{+}\tau^{-}$ ($b \bar{b}$) with cross sections of 
$\langle \sigma v \rangle \sim 10^{-26}\ {\rm cm}^{3} \ {\rm s}^{-1}$ 
\citep{Goodenough:2009gk,Hooper:2010mq,Boyarsky:2010dr,
Hooper:2011ti,Abazajian:2012pn,Gordon:2013vta,Abazajian:2014fta,Daylan:2014rsa}.
The cross-correlation signal offers an independent method for testing the DM interpretation
of the excess.

\begin{table}[!t]
\begin{center}
\begin{tabular}{|c|c|c|c|c|}
\tableline
& ev2/P7V6 & ev2/P7rep & ev4/P7V6 & ev4/P7rep  \\ \tableline
W1 &6.91/10    & 6.22/10   & 8.58/10    & 7.80/10\\ \tableline
W2 &12.26/10  & 12.32/10  & 6.98/10   & 6.87/10\\ \tableline
W3 &7.62/10    & 7.11/10    & 8.77/10   & 6.49/10\\ \tableline
W4 &12.88/10  & 12.95/10  & 7.57/10    & 7.39/10\\ \tableline
\end{tabular} 
\caption{
\label{tab:check_fermi_ana}
The impact of $Fermi$ Galactic diffuse model on the cross-correlation analysis.
We summarize the $\chi^2$ value of the cross-correlation signal 
in each CFHTLenS patch using different models and photon selections.
}
\end{center}
\end{table}

Encouraged by our initial study producing competitive constraints, we investigate
the improvement expected with 
upcoming gravitational lensing survey with the sky coverage of 20000 
square degs. 
We have shown that constrints on $\langle \sigma v \rangle$ 
would reach $2.7-22.2\times10^{-26} \ {\rm cm}^{3} \ {\rm s}^{-1}$ for the $b \bar{b}$ 
channel and $1.1-8.51\times10^{-25} \ {\rm cm}^{3} \ {\rm s}^{-1}$ for the $\tau^{+}\tau^{-}$ 
channel, both for a 100 GeV DM. For lighter DM motivated by the Galactic center excess,
the constraints would reach 
$1.34-10.96\times10^{-26} \ {\rm cm}^{3} \ {\rm s}^{-1}$ for the $b \bar{b}$ 
channel (assuming $40$ GeV mass) 
and $0.39-3.24\times10^{-26} \ {\rm cm}^{3} \ {\rm s}^{-1}$ 
for the $\tau^{+}\tau^{-}$ (assuming $10$ GeV mass), allowing a test of the DM origin of the 
Galactic center excess.
Furthermore, if the 
accurate modeling of astrophysical contributions to the cross-correlation can be made, 
one can reasonably expect constraints on $\langle \sigma v \rangle$ to improve by 40-70\% 
for a broad range of DM mass.
Gamma-ray data also stand to improve. In this study we have used a conservative 
mask of $2^\circ$ around each point-source. While more aggressive masks or point-source
modeling will increase photon statistics, these must be weighed by their larger systematic
uncertainties. Also, at present, when we adopt a smaller mask of $1^\circ$ radius around each 
point source, we find that the errors on $\xi$ improved by only 10\%. Nevertheless, with more 
data, aggressive masks will become feasible. In particular, analyses that focus on higher energy 
photons, which due to their higher angular and energy resolutions can tolerate more aggressive 
masks, may yield improved probes especially at high DM masses.

Overall, these results suggest that the cross-correlation analysis of cosmic shear and the EGB
will play a crucial role for the search for DM annihilation signatures. It is thus
important to address a few issues in the cross-correlation analysis of cosmic shear 
and the EGB. First, in this paper, we have only implemented a crude estimate of the 
systematic error associated with the gamma-ray foreground subtraction. 
Second, we have not included the sampling variance. 
While these are not expected to be a significant source of uncertainties
at present, mainly because of the large statistical error in the current 
data sets, they would become more important for analyses using data from 
upcoming surveys. For the diffuse model 
subtraction, we have made an attempt to estimate the systematics 
by employing different gamma-ray datasets and different 
Galactic diffuse emission models. The resulting $\chi^2$ values in each of the CFHTLenS 
patches are summarized in Table~\ref{tab:check_fermi_ana}, and shows how the typical 
systematic error associated with {\it Fermi} photon analysis are very small ($\Delta \chi^2 \sim 1$--$5$). 
In the case of a LSST-like survey (see Section~\ref{subsec:forecast}), this difference could 
induce a systematic error of $\langle \sigma v\rangle$ for a 100 GeV DM on the level of
$\sim 3\times10^{-26} \ {\rm cm}^{3} \ {\rm s}^{-1}$ 
for both the $b \bar{b}$ channel and
the $\tau^{+}\tau^{-}$ channel. 

Detailed comparisons with numerical simulations would also be needed to test the 
accuracy of our benchmark model based on halo model approach (see also Appendix B).
Combined with other observabations such as the mean intensity of the EGB, angular correlation
of the EGB and the cross-correlation of galaxy position and the EGB \citep{Xia:2011ax}, 
one can expect that some of the degeneracies between the DM annihilation and 
astrophysical sources may be broken. 
It is therefore important to investigate how much information of the EGB can be extracted 
from such combined analyses using multiple astrophysical datasets.
Gamma-ray analyses with future cosmological surveys would be very powerful methods 
for understanding the origin of the EGB and the indirect search of DM annihilation.

\begin{acknowledgements}
The authors are grateful to Kazuhiro Nakazawa for helpful discussion on cross correlation analysis.
MS is supported by Research Fellowships of the Japan Society for 
the Promotion of Science (JSPS) for Young Scientists, and
SH is supported by a Research Fellowship for Research Abroad by JSPS. 
NY acknowledges financial support from
the JSPS
Grant-in-Aid for Scientific Research (25287050). 
Numerical computations presented in this paper were in part carried out
on the general-purpose PC farm at Center for Computational Astrophysics,
CfCA, of National Astronomical Observatory of Japan.
This work is based on observations obtained with MegaPrime/MegaCam, 
a joint project of CFHT and CEA/IRFU, at the Canada-France-Hawaii Telescope 
(CFHT) which is operated by the National Research Council (NRC) of Canada, 
the Institut National des Sciences de l'Univers of the Centre National 
de la Recherche Scientifique (CNRS) of France, and the University of Hawaii. 
The research used the facilities of the Canadian Astronomy Data Centre 
operated by the National Research Council of Canada with the support of 
the Canadian Space Agency. CFHTLenS data processing was made possible 
thanks to significant computing support from the NSERC Research Tools 
and Instruments grant program.
\end{acknowledgements}

\appendix
\section{\label{sec:appendixA}Covariance of Cross-Correlation Estimator}

Here, we summarize the properties of the estimator for cross-correlation 
analysis used in the present paper. Our estimator is given by Eq.~(\ref{eq:CCest}).
Let us consider a simple case in this appendix.
When one measures galaxies' ellipticities (${\bd \epsilon}$) 
and counts extragalactic gamma-ray photons ($\delta n$) 
from an observed data set precisely,
the cross-correlation estimator is expressed by
\beqa
{\hat \xi}_{\delta n-\gamma_{t}}(\theta)
&=&\frac{1}{N_{\rm p}(\theta)}
\sum^{N_{\rm pixel}}_{i}\sum^{N_{\rm gal}}_{j}
\delta n({\bd \phi}_{i})\epsilon_{t}({\bd \phi}_{j}|{\bd \phi}_{i})
\Delta_{\theta}({\bd \phi}_{i}-{\bd \phi}_{j}), \label{eq:CCest_simple} \\
N_{\rm p}(\theta) 
&=&
{\displaystyle \sum^{N_{\rm pixel}}_{i}\sum^{N_{\rm gal}}_{j}
\Delta_{\theta}({\bd \phi}_{i}-{\bd \phi}_{j})},
\eeqa
where $\Delta_{\theta}({\bd \phi}) = 1$
for $\theta-\Delta \theta/2 \le \phi \le \theta+\Delta \theta/2$ and zero otherwise
and $N_{\rm p}(\theta)$ represents the effective pair number in cross-correlation analysis.
One can clearly see that this estimator is an unbiased estimator of
of cross-correlation signal $\xi_{\delta n-\gamma_{t}}(\theta)$.

In order to discuss statistical significances of the measured estimator from real data, 
we need to estimate the covariance of ${\hat \xi}_{\delta n-\gamma_{t}}(\theta)$.
In particular, the covariance 
in the case of $\langle {\hat \xi}_{\delta n-\gamma_{t}}(\theta)\rangle =0$
is needed for detection of cross-correlation signals.
The covariance matrix of Eq.~(\ref{eq:CCest_simple}) is defined by
\beqa
{\rm Cov} \left[{\hat \xi}_{\delta n-\gamma_{t}}(\theta_{1}), {\hat \xi}_{\delta n-\gamma_{t}}(\theta_{2}) \right] &=&
\langle 
({\hat \xi}_{\delta n-\gamma_{t}}(\theta_{1}) -\xi_{\delta n-\gamma_{t}}(\theta_{1}))
({\hat \xi}_{\delta n-\gamma_{t}}(\theta_{2}) -\xi_{\delta n-\gamma_{t}}(\theta_{2}))
\rangle \nonumber \\
&=& \frac{1}{N_{\rm p}(\theta_1)N_{\rm p}(\theta_2)}
\Bigg[\sum_{i, j, k, \ell}
\langle
n({\bd \phi}_{i})\epsilon_{t}({\bd \phi}_{j}|{\bd \phi}_{i})
n({\bd \phi}_{k})\epsilon_{t}({\bd \phi}_{\ell}|{\bd \phi}_{k})
\rangle 
\nonumber \\ 
&&\, \, \, \, \,
\times \Delta_{\theta_1}({\bd \phi}_{i}-{\bd \phi}_{j})
\Delta_{\theta_2}({\bd \phi}_{k}-{\bd \phi}_{\ell})\Bigg]
-\xi_{\delta n-\gamma_{t}}(\theta_{1})\xi_{\delta n-\gamma_{t}}(\theta_{2}),
\eeqa
where $i$ and $k$ represents the indeces of summation over gamma-ray counts,
and $j$ and $\ell$ are for galaxies.
When two fields $\delta n$ and ${\bd \epsilon}$ are independent of each other,
the ensemble average $\langle \delta n \, \epsilon_{t} \, \delta n \, \epsilon_{t} \rangle$
would simply reduce the ensemble average of each field, i.e.
$\langle \delta n \, \delta n\rangle \langle \epsilon_{t} \, \epsilon_{t} \rangle$.
For shape of galaxies, the two point correlation function 
$\langle \epsilon_{t} \, \epsilon_{t} \rangle$ would be expressed by the summation 
of intrinsic variance and the correlation signal due to large scale structure;
\beqa
\langle \epsilon_{t}({\bd \phi}_{j})\epsilon_{t}({\bd \phi}_{\ell})\rangle
= \frac{\sigma_{\rm int}^2}{2} \delta_{j\ell}+\xi_{+} (|{\bd \phi}_{j}-{\bd \phi}_{\ell}|),
\label{eq:2pcf_shear}
\eeqa
where $\sigma_{\rm int}$ represents the variance of intrinsic shape of galaxies
and $\xi_{+}(\theta)$ is the two point correlation signal due to weak gravitational lensing. 
In a concordance $\Lambda$CDM universe, $\xi_{+}(\theta)$ would 
be expected to be on the order of $10^{-4}$.
The latest cosmic shear measurement \citep{Kilbinger:2012qz} confirmed this expectation 
with high significance and shows that the typical value of $\sigma_{\rm int}$ to be $\sim$ 0.4.
For extragalactic gamma-ray counts, the origin is still unknown.
Hence, it is difficult to estimate the exact contribution to the two point correlation function
$\langle \delta n \, \delta n \rangle$.
At least, we expect that Poisson processes would 
dominate on scales larger than the PDF in gamma-ray surveys.
We assume that photon count fluctuations follow a Poisson distribution 
with mean corresponding to $\delta n^{\rm obs}({\bd \phi})$,
where $\delta n^{\rm obs}({\bd \phi})$ is the observed gamma-ray count map.
In this case, two point correlation function 
$\langle \delta n \, \delta n\rangle$ would be expressed by
\beqa
\langle \delta n({\bd \phi}_{i})\delta n({\bd \phi}_{k})\rangle
= \delta n^{\rm obs}({\bd \phi}_{i})\delta_{ik}
+ \delta n^{\rm obs}({\bd \phi}_{i})\delta n^{\rm obs}({\bd \phi}_{k}), 
\label{eq:2pcf_egb}
\eeqa
where the first term represents Poisson fluctuations in count maps
and the second term includes the effect of correlation due to 
the point spread function in gamma-ray surveys.
Eq.~(\ref{eq:2pcf_egb}) would be a reasonable approximation
when considering scales larger than the size of point spread function, 
i.e. $\sim 1$ deg in our analysis.

Using Eqs.~(\ref{eq:2pcf_shear}) and (\ref{eq:2pcf_egb}),
and $\langle {\hat \xi}_{\delta n-\gamma_{t}}(\theta)\rangle =0$, 
we can divide the covariance of our estimator into four contributions as follows:
\beqa
{\rm Cov} \left[{\hat \xi}_{\delta n-\gamma_{t}}(\theta_{1}), {\hat \xi}_{\delta n-\gamma_{t}}(\theta_{2}) \right] &=&
C_{\rm SN+p}(\theta_1, \theta_2)
+C_{\rm WL+p}(\theta_1, \theta_2) \nonumber \\
&&
\, \, \, \, \, \, \, \, \, \, \, \, \, \, \, \, \, \, \, \, \, \, \, \, \, \, \,
+C_{\rm SN+obs}(\theta_1, \theta_2)
+C_{\rm WL+obs}(\theta_1, \theta_2), 
\label{eq:cov_est_full}
\\
C_{\rm SN+p}(\theta_1, \theta_2) &=&
\frac{1}{N_{\rm p}(\theta_1)N_{\rm p}(\theta_2)}
\sum_{i,j}\delta n^{\rm obs}({\bd \phi}_{i})\frac{\sigma_{\rm int}^2}{2}
\Delta_{\theta_{1}}(ij)
\Delta_{\theta_{2}}(ij), \\
C_{\rm WL+p}(\theta_1, \theta_2) &=&
\frac{1}{N_{\rm p}(\theta_1)N_{\rm p}(\theta_2)}
\sum_{i,j,\ell}\delta n^{\rm obs}({\bd \phi}_{i})\xi_{+}(|{\bd \phi}_{j}-{\bd \phi}_{\ell}|)
\Delta_{\theta_{1}}(ij)
\Delta_{\theta_{2}}(i\ell), \\
C_{\rm SN+obs}(\theta_1, \theta_2) &=&
\frac{1}{N_{\rm p}(\theta_1)N_{\rm p}(\theta_2)}
\sum_{i,j,k}
\delta n^{\rm obs}({\bd \phi}_{i})\delta n^{\rm obs}({\bd \phi}_{k})
\frac{\sigma_{\rm int}^2}{2}
\Delta_{\theta_{1}}(ij)
\Delta_{\theta_{2}}(kj), \\
C_{\rm WL+obs}(\theta_1, \theta_2) &=&
\frac{1}{N_{\rm p}(\theta_1)N_{\rm p}(\theta_2)}
\Bigg[
\sum_{i,j,k,\ell}\delta n^{\rm obs}({\bd \phi}_{i})n^{\rm obs}({\bd \phi}_{k})
\xi_{+}(|{\bd \phi}_{j}-{\bd \phi}_{\ell}|) \nonumber \\
&&
\,\,\,\,\,\,\,\,\,\,\,\,\,\,\,\,\,\,\,\,\,\,\,\,\,\,\,\,
\,\,\,\,\,\,\,\,\,\,\,\,\,\,\,\,\,\,\,\,\,\,\,\,\,\,\,\,
\,\,\,\,\,\,\,\,\,\,\,\,\,\,\,\,\,\,\,\,\,\,\,\,\,\,\,\,
\times \Delta_{\theta_{1}}(ij)
\Delta_{\theta_{2}}(k\ell)
\Bigg],
\eeqa
where $\Delta_{\theta_1}(ij) = \Delta_{\theta_1}({\bd \phi}_{i}-{\bd \phi}_{j})$ and so on.
According to the observational fact that $\xi_{+}$ 
is smaller than $\sigma_{\rm int}^{2}$ by a factor of $10^{-3}$, 
the dominant contributions in Eq.~(\ref{eq:cov_est_full}) would be
the first term $C_{\rm SN+p}$ and 
the third term $C_{\rm SN+obs}$.
$C_{\rm SN+p}$ is estimated from the observed galaxy catalogue
and random count maps based on Poisson distribution.
We can also estimate $C_{\rm SN+obs}$ by cross-correlating
the observed photon counts and randomized galaxy catalogues.
The estimation of $C_{\rm SN+p}$ and $C_{\rm SN+obs}$ from the real data set
is found in Section \ref{subset:analysis}.

\section{\label{sec:appendixB}Effect Of Dark Matter Halo Profile Uncertainties On Cross-Correlation Signals}

Here, we quantity the effect of uncertainties of the DM halo profiles on the cross-
correlation between cosmic shear and the EGB.
In order to calculate the theoretical model of cross-correlation signals, 
we follow the halo model approach as in Section \ref{subsec:halomodel}.
The halo model posits that there are mainly two contributions of the cross-correlation 
signal: the one-halo term and
the two-halo term. For a given length scale $k$, the main contribution to the one-halo term 
as calculated by Eq.~(\ref{eq:1h}) comes from galaxy cluster size halos with $10^{13}-10^{15} M_{\odot}$.
This is valid for the two-halo term associated with 
density fluctuations (i.e., the first integral in 
Eq.~(\ref{eq:2h})). On the other hand, the two-halo term associated with density squared (i.e.,
the second integral in Eq.~(\ref{eq:2h}))
is mainly determined by the smoothed profile contribution
$\int {\rm d}V \rho^2_{h}(r|M, z)$ with dominant contribution from lower mass scales.
Assuming that the concentration parameter $c_{\rm vir} = r_{\rm vir}/r_{s} \propto M^{\alpha}$ with $\alpha \sim -0.1$,
$M n(M,z)\int {\rm d}V \rho^2_{h}(r|M, z)$ would scaled as $\sim M^{3\alpha}$ for $M < 10^{12} M_{\odot}$.
This fact implies that the low mass halos dominates the two-halo term and that
the overall amplitude of the two-halo term is sensitive to the minimum halo mass. 
Thus, along with $M_{min}$, $c_{\rm vir}(z, M)$ 
is one of the most important parameters in the halo model.

\begin{figure}[!t]
\begin{center}
       \includegraphics[clip, width=0.45\columnwidth]{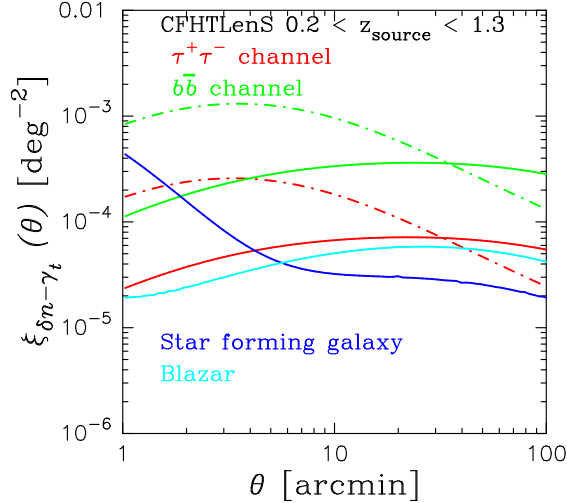}
       
    \caption{
    \label{fig:xi_dm_prada+12}
   The expected cross-correlation signals of cosmic shear and EGB from various sources. 
   The signal from the annihilation of a 100 GeV 
   mass DM particle with annihilation cross section $\langle \sigma v \rangle=3 \times 10^{-26} \ {\rm cm}^3 \ {\rm s}^{-1}$
   is shown separately for the $\tau^{+}\tau^{-}$ channel (red lines) and the $b {\bar b}$ channel (green lines).
   The solid lines shows the halo model with the power-low model of $c_{\rm vir}$
   with assumed minimum DM halo mass $M_{\rm min} = 10^{-6} M_\odot$.
   The dashed-dotted line corresponds to the halo model calculation with the non-monotonic model of 
   $c_{\rm vir}$.
   The blue and cyan line show the 
   contribution from SFG and blazars, respectively.
  } 
    \end{center}
\end{figure}

Recent numerical simulations \citep[e.g.,][]{Prada:2011jf} suggest a non-monotonic relation between 
the concentration parameter and the mass of DM haloes. In this appendix, we test the dependence of 
the cross-correlation signal on $c_{\rm vir}(z, M)$ by comparing a simple power-law model and the 
non-monotonic model.
For the non-monotonic $c_{\rm vir}(z, M)$ model, we use the fitting function of Ref. \citep{Prada:2011jf} that determines $c_{\rm vir}$ as a function of the linear $rms$ density fluctuation $\sigma(z, M)$.
This fitting function successfully reproduces the complex feature of $c_{\rm vir}$ found in 
numerical simulations. For the power-law model, we apply the functional form shown in 
Ref.~\citep{Bullock:1999he} as in our benchmark model.

\begin{figure}[!t]
\begin{center}
       \includegraphics[clip, width=0.45\columnwidth]{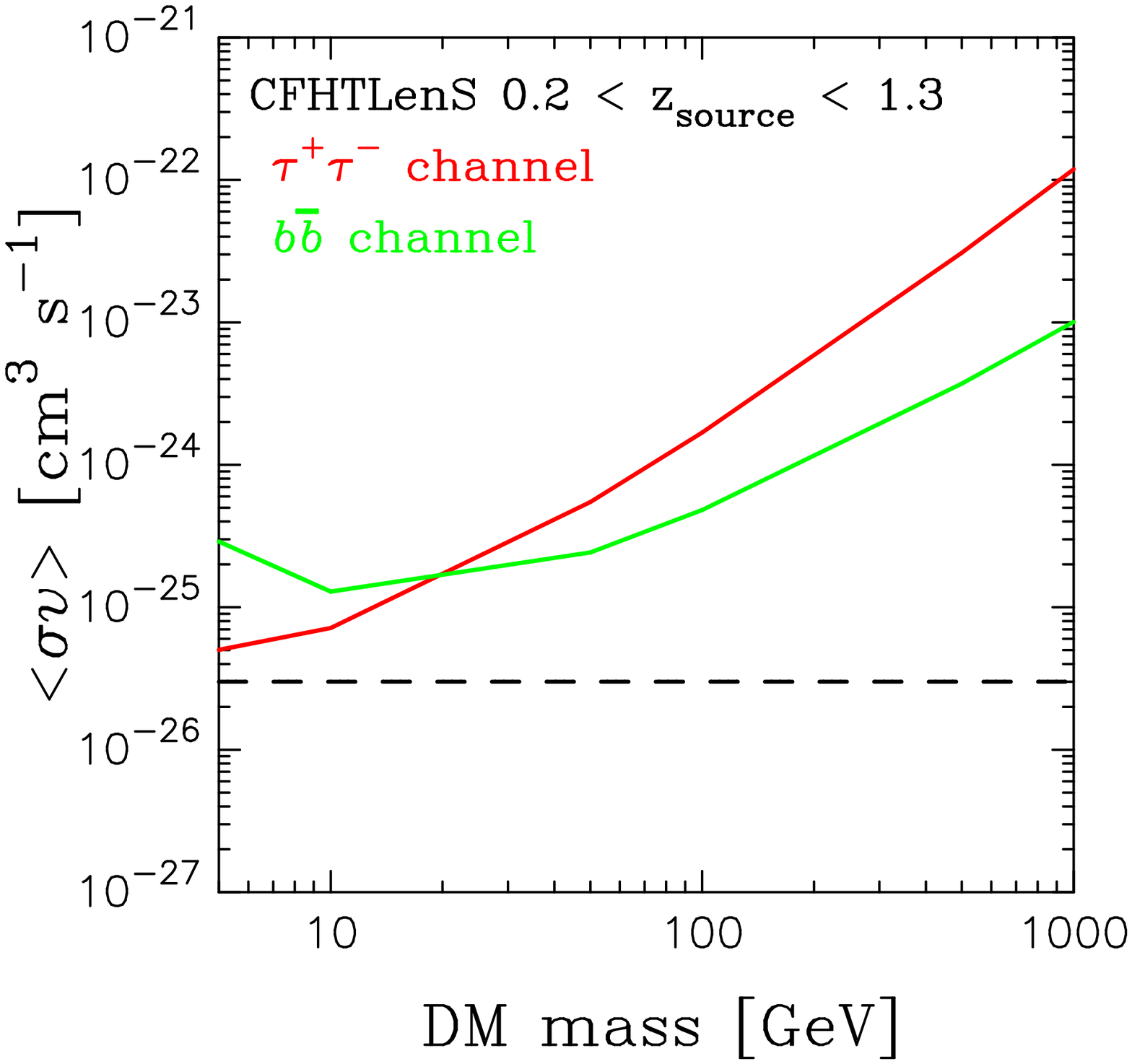}
    \caption{
     \label{fig:sigv_m_CFHT_prada+12}
    The 68 \% confidence level upper limits on $\langle \sigma v \rangle$ 
   as functions of the DM mass.
   The red shaded region shows the upper limit for the $\tau^{+}\tau^{-}$ channel
   and the green one for the $b {\bar b}$ channel.
	     } 
    \end{center}
\end{figure}

Figure \ref{fig:xi_dm_prada+12} shows the comparison between the halo model calculations 
with the power-law and non-monotonic models of $c_{\rm vir}$.
Each solid line is the same as the one shown in figure \ref{fig:xi_dm}.
The dashed-dotted lines correspond to the halo model with the non-monotonic model of $c_{\rm vir}$.
For the non-monotonic model of $c_{\rm vir}$, we found that the final result is much less sensitive 
to the minimum halo mass because of the flattening feature of $c_{\rm vir}$ at low masses.
The most important result is perhaps that 
the cross-correlation signals would be dominated by the one-halo term for the non-monotonic model,
which is different from the result of our benchmark model and 
from previous work \citep{Camera:2012cj}.
This is mainly due to the higher concentration in massive DM haloes than in our benchmark model.
Consequently, 
the expected signals for the non-monotonic model 
would be ten times as large as our benchmark model for smaller angular scale at $\theta < 10$ arcmin.
However, for the angular scale larger than 30 arcmin, 
the two models with the different $c_{\rm vir}$ show quite similar amplitudes of the cross-correlation.
Clearly, the choice of $c_{\rm vir}$ model would not 
affect the final constraints of DM annihilation significantly
because most of the information about DM annihilation come from the large scale clustering 
as shown in Section~\ref{subsec:forecast}.
Figure \ref{fig:sigv_m_CFHT_prada+12} shows 
the 68 \% confidence upper limit of DM annihilation 
obtained from the current data set shown in Section~\ref{sec:data}
with the non-monotonic model of $c_{\rm vir}$.
In figure \ref{fig:sigv_m_CFHT_prada+12}, 
we simply assume that DM annihilation is the only contribution to
the cross-correlation signals
and take into account the smoothing effect due to PSF in the same manner shown 
in Section~\ref{subsec:halomodel}.
We found the constraints on $\langle \sigma v \rangle$ degrade by $\sim$ 10 \%
over a wide mass range of 5--1000 GeV.

\bibliography{ref_prd}
\end{document}